\documentclass[12pt]{article}

\usepackage[utf8]{inputenc}
\usepackage[english]{babel}

\usepackage{authblk} 
\usepackage{titlesec}

\titleformat{\section}
  {\normalfont\fontsize{15}{15}\bfseries}{\thesection}{1em}{}
\titleformat{\subsection}
  {\normalfont\fontsize{13}{15}\bfseries}{\thesubsection}{1em}{}

\usepackage{amssymb, amsmath, bm, mathtools}
\usepackage[displaymath]{lineno}
\usepackage{gensymb}
\usepackage{siunitx}
\usepackage{graphicx}
\usepackage{pgf}
\usepackage{pgfpages}
\usepackage{tikz}
\usepackage{ifthen}
\usepackage{hyperref}
\usepackage{cleveref}
\usepackage{multirow}
\usepackage{algorithm}
\usepackage[noend]{algpseudocode}
\usepackage{adjustbox}
\usepackage{comment}
\usepackage[a4paper, left=2.5cm, right=2.5cm, top=3cm, bottom=3cm]{geometry}

\usepackage{xcolor} %
\usepackage{enumerate} %

\usepackage[version=4]{mhchem}
\usepackage{nccmath}
\usepackage{wrapfig}

\usepackage{accents}

\usepackage{xspace}

\usetikzlibrary{fit}
\usetikzlibrary{angles}
\usepackage{setspace}
\usepackage{soul}

\usepackage{xr}

\makeatletter
\def\BState{\State\hskip-\ALG@thistlm}
\makeatother %

\newcommand*{\dt}[1]{%
  \accentset{\mbox{\large\bfseries .}}{#1}}

\newcommand{\ie}{{\textit{i.e.}}}
\renewcommand{\vector}[1]{\boldsymbol{#1}}%
\newcommand{\pH}{\mathrm{pH}}
\newcommand{\kinetic}{\kappa}

\newcommand{\ox}{\ce{\kappa_{ox}}}
\newcommand{\red}{\ce{\kappa_{red}}}
\newcommand{\metal}{\mathrm{M}}
\newcommand{\astate}{\mathtt{a}}
\newcommand{\bstate}{\mathtt{b}}

\newcommand{\ferrous}{\ce{Fe^{II}}}
\newcommand{\ferric}{\ce{Fe^{III}}}

\usepackage{tocbibind} %
\usepackage[toc,page]{appendix} %
\usepackage{adjustbox}
\usepackage{bbold}
\modulolinenumbers[1]
\graphicspath{{./}}

\usepackage[numbers,sort&compress]{natbib}
\begin{document}
\title{Navigating the Complexities of Multiple Redox State Interactions in Aqueous Systems}
\author[1]{Shishir Mundra \thanks{Corresponding Author: shishir.mundra@ifb.baug.ethz.ch} } 
\author[1]{Mohit Pundir}
\author[2]{Barbara Lothenbach}
\author[1]{David S. Kammer}
\author[1]{Ueli M. Angst}

\affil[1]{Institute for Building Materials (IfB), ETH Zürich, Laura-Hezner-Weg 7, 8093, Zürich, Switzerland}
\affil[2]{EMPA, Laboratory for Concrete \& Asphalt, CH-8600 Dübendorf, Switzerland} \date{}
\maketitle

\section*{Abstract}
Numerous aqueous systems host elements in multiple redox states, with wide ranging implications such as their influence on the formation/dissolution of minerals, water toxicity, and nutrient cycling. To uncover governing mechanisms and complex chemical interactions in aqueous systems, reactive-transport models have increasingly gained importance. However, their predictive capabilities remain limited because existing approaches struggle to accurately account for the full complexities of redox reactions. Here, we develop a reactive-transport framework that leverages recent advancements in thermodynamic modelling, speciation chemistry, and redox kinetics. Distinct from traditional models, we uniquely treat redox kinetics along with transport as transient phenomena, decoupled from Gibbs free energy minimisation. Ensuring ion concentrations are governed by non-equilibrium rate laws, this approach allows predicting the tempo-spatial distribution of speciation and precipitation of species across oxidation states. We illustrate the versatility of our framework through two case studies: manganese speciation in natural waters and the fate of dissolved iron in aqueous/porous media. Our framework significantly enhances the modelling of a diverse range of redox-sensitive environments. 

\noindent \textbf{Keywords:} Redox reactions; Reactive-transport framework; Oxidation states; Thermodynamic speciation; Kinetics

\newpage

\section{Main}

Several elements (such as Fe, Mn, Cr) can co-exist in multiple redox states in natural and man-made aqueous environments. The activity of an element in its various redox states influences both the thermodynamic state of the aqueous system and processes like the precipitation or dissolution of solid phases, as well as the transport of aqueous species \cite{Stumm1996AquaticWaters}. Across multiple oxidation states, the activity of an element is kinetically controlled by the local environmental conditions, such as pH, temperature, \ce{pO_2}, concentration of oxidants and reductants, complexation, adsorption or nucleation sites \cite{Stumm1996AquaticWaters}. For instance, the oxidation of dissolved \ferrous{} to \ferric{} in different aqueous and porous environments proceeds through multiple simultaneous oxidation reactions depending on the pH and the prevalent Fe-species~\cite{Mundra2023AerobicSolutions, Stumm1961OxygenationIron, Pham2008OxygenationPathways, Millero1987OxidationStrength, Weiss1935ElektronenubergangsprozesseLosungen}. Similar complex reaction schemes exist for other elements \cite{Sharma1988EffectWaters, Millero1985TheWaters} such as the oxidation and reduction reactions involving aqueous \ce{Mn^{II}}, \ce{Mn^{III}} or \ce{Mn^{IV}} in oceanic water \cite{Oldham2020TheOcean, Trouwborst2006SolubleZones, Klewicki1996TheComplexes, Morgan2005KineticsSolutions} and \ce{Cr^{III}}, \ce{Cr^{IV}}, \ce{Cr^{V}} and \ce{Cr^{VI}} in groundwater \cite{Zhitkovich2011ChromiumRisks, Varadharajan2017ReoxidationRegimes}. To understand the impact of these elements' oxidation states on their environment and its processes, it is essential to accurately predict their spatial and temporal evolution. Predicting such complex reactions, involving multiple species, kinetically controlled conversion, and transport, poses challenges and often necessitates advanced modelling techniques \cite{Lensing1994ModelingSubsurface, Marzal1994ModelingConditions}. 

Reactive transport models couple the reaction process with the transport process to simulate evolution of chemical species across space and time \cite{Steefel2019ReactiveCrossroads}. Most reactive transport models solve the coupled equations using an operator split approach, where, in each iteration step, the conservative transport equations are solved first, followed by solving the chemical reaction terms using the transported concentrations \cite{damiani_framework_2020, Sun2019}. This method facilitates the use of advanced thermodynamic solvers like GEM-Selektor~\cite{kulik_gem-selektor_2012, Kulik2013GEM-SelektorCodes}, Reaktoro~\cite{leal_computational_2016}, and PHREEQC~\cite{Parkhurst2013DescriptionCalculations} to handle complex chemical reactions either in total equilibrium or in partial equilibrium for kinetically-driven reactions~\cite{VanCappellen1996CyclingManganese}. However, their usage necessitates accurate datasets on the equilibrium constants for each of the species involved in the chemical system~\cite{VanCappellen1996CyclingManganese,Hummel2023The2020}. While this approach has been effective for various chemical environments \cite{leal_overview_2017}, it presents challenges, particularly when considering kinetically controlled multiple redox reactions. Oxidation and reduction reactions are constituted of several reaction pathways involving the speciation of elements in different oxidation states~\cite{Mundra2023AerobicSolutions, Sharma1988EffectWaters, Millero1985TheWaters,Oldham2020TheOcean, Trouwborst2006SolubleZones, Klewicki1996TheComplexes, Morgan2005KineticsSolutions,Zhitkovich2011ChromiumRisks, Varadharajan2017ReoxidationRegimes}, each with a different kinetic constant. Solving such kinetically-driven chemical systems in partial equilibrium requires accurate knowledge of the kinetic constants and the molar fractions of each of the species involved in the redox reaction and its pathways. However, numerical estimates of the kinetic rate constants for each redox reaction pathway de-convoluted from the experimentally derived rate law commonly reported, vary by several orders of magnitude \cite{Millero1987OxidationStrength, Pham2008OxygenationPathways, Santana-Casiano2005OxidationWaters, Santana-Casiano2004TheModel, WhitneyKing1998RoleSystems, morgan_effect_2007}. Such large variations, compounded with incomplete thermodynamic datasets on aqueous speciation, has inhibited partial equilibrium solvers to accurately predict the fate of elements in different redox states and their consequences on dissolution or precipitation behaviour.

To overcome these challenges associated with modelling of chemical systems that involve elements in multiple redox states, we introduce a novel reactive transport framework. We employ a staggered approach, where we first consider kinetic rate laws governing redox reactions along with other transient phenomena such as transport. Unlike partial equilibrium approaches where the entire chemical system including all oxidation states is solved in a monolithic manner, here we decouple the Gibbs free energy minimisation for deciphering the speciation and precipitation of elements in different oxidations states. We take advantage of the availability of accurate thermodynamic datasets to iteratively solve the decoupled systems. Furthermore, this decoupling enables the use of kinetic constants of overall redox reactions, for which reliable experimental data is available~\cite{Mundra2023AerobicSolutions,Morgan2005KineticsSolutions, Klewicki1996TheComplexes}. This allows our reactive transport framework to simultaneously predict the fate of elements in multiple redox states (thermodynamically stable and metastable species) and its influence on the evolution of the overall chemistry of the aqueous system both in space and time. 

\begin{figure}[!ht]
    \centering
    \includegraphics[width= 0.95\textwidth]{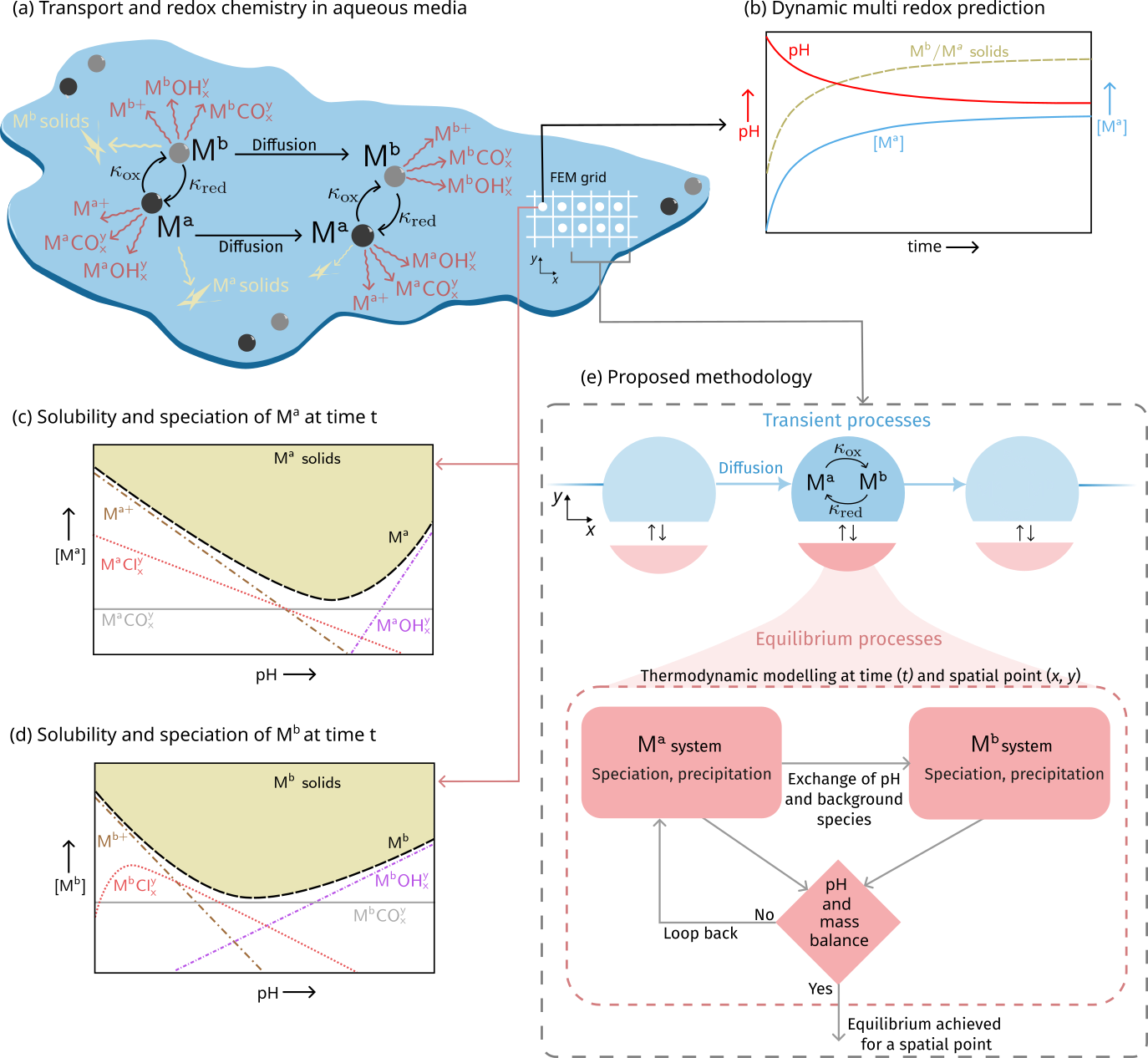}
    \caption{\textbf{Conceptual modelling approach for solving the spatial and temporal evolution of multi-redox species in the proposed reactive transport framework:} (a) Schematic of the complex interplay between various kinetically controlled processes such as speciation, redox reactions, precipitation and transport, governing the fate of element ($\metal$) in multiple redox states ($\metal^\astate$ and $\metal^\bstate$, where $\astate$ $>$ 0, $\bstate$ $>$ $\astate$, and $\astate$ and $\bstate$ are positive integers) in aqueous media. (b) The temporal evolution of multi-redox species ($\metal^\astate$ and $\metal^\bstate$) at a certain location in space is governed by ion transport and kinetically controlled redox reactions that depend on electrolyte chemistry, and (c)-(d) pH-dependent aqueous speciation and solubility of $\metal^\astate$ and $\metal^\bstate$. (e) Our proposed methodology navigates such complexities of redox sensitive aqueous environments by decoupling transient (blue) and equilibrium processes (red). The proposed method solves for hydrolysis, speciation, oxidation or reduction, precipitation, dissolution, transport of elements (in multiple redox states).
    }.
    \label{fig:methodology}
\end{figure}

\section{Fundamental concept behind the proposed reactive-transport framework}
We propose a reactive-transport framework that incorporates the key mechanisms affecting a redox chemical system, such as hydrolysis and complexation, precipitation, kinetics of redox reactions, and the transport of aqueous and solid species (\Cref{fig:methodology}a). We differentiate the above mechanisms as either transient (time-dependent) or equilibrium processes (indicated in blue and red, respectively in \Cref{fig:methodology}e). A unique feature in our approach is that we consider the kinetically-controlled oxidation and reduction of metal ions, along with the transport of all aqueous and solid species, as transient processes (\Cref{fig:methodology}e). By considering the rate laws governing redox reactions as transient processes, thus separately from thermodynamic modelling, we ensure that the concentrations of metal ions in different oxidation states are determined by non-equilibrium processes \cite{Lindberg1984GroundEh} (\Cref{fig:methodology}b). Numerically, we implement this by introducing variables \ce{[M^{$\astate$}]} and \ce{[M^{$\bstate$}]} for aqueous ions in each redox state, formed either because of oxidation or reduction. Here, \ce{[M^{$\astate$}]} and \ce{[M^{$\bstate$}]} denote the sum of the concentration of all aqueous complexes of an element M in oxidation state $\astate$ and $\bstate$, respectively (\Cref{fig:methodology}a, e). This approach enables us to use rate constants of oxidation (\ox) or reduction (\red), that can be experimentally determined and for which generally reliable data is available in the literature~\cite{Pham2008OxygenationPathways,Mundra2023AerobicSolutions}. Consequently, we tackle the challenges pertaining to the limited availability of kinetic rate constants for each of the species participating in redox reactions (\Cref{fig:supp-fe}b). The spatial and temporal evolution of \ce{[M^{$\astate$}]} depends not only on its diffusion but also on its rate of conversion (oxidation or reduction) to \ce{[M^{$\bstate$}]} (shown within blue regions in \Cref{fig:methodology}e). Within each time increment, we first address the transient phenomena for each aqueous species, see (\Cref{sec:method} for more details) and then utilise the acquired concentration values to address the equilibrium phenomena (red regions in \Cref{fig:methodology}e).

After solving the transient processes, we utilise the variables \ce{[M^{$\astate$}]} and \ce{[M^{$\bstate$}]} to solve for the speciation and precipitation of ions in multiple redox state at every spatial point (\Cref{fig:methodology}c, d). Here, we separate speciation and precipitation of one oxidation state from the speciation and precipitation of the other oxidation state (represented by red in \Cref{fig:methodology}e). For this, we take the \ce{[M^{$\astate$}]} corresponding to oxidation state $\astate$ at a spatial point, obtained after diffusion and oxidation/reduction from the first stage (indicated in blue in \Cref{fig:methodology}e), along with the remaining amount of background species. We solve this chemical system under thermodynamic equilibrium (detailed in \Cref{sec:method}) under varying environmental conditions such as pH, partial pressures of \ce{CO_2} and \ce{O_2}, and temperatures. Next, we consider a separate chemical system that comprises the \ce{[M^{$\bstate$}]} corresponding to a different oxidation state, $\bstate$, (obtained from the oxidation or reduction considered as transient processes) and the remaining background species (obtained from the previous chemical system). We solve the chemical systems corresponding to different oxidation states in parallel and in an iterative manner where all the background species are constantly exchanged between the systems, ensuring they converge to the same pH value (see \Cref{fig:methodology}) and are mass balanced. The concentration of different species determined in this way (including multiple redox complexes, background species, and pH) reflects the concentration at the end of the time interval and serves as the starting point for the subsequent time step. Please refer to \Cref{sec:method} for a detailed description of the methodology.

\section{Application of the reactive-transport framework}

\begin{figure}[H]
    \centering
    \includegraphics[width=0.95\textwidth]{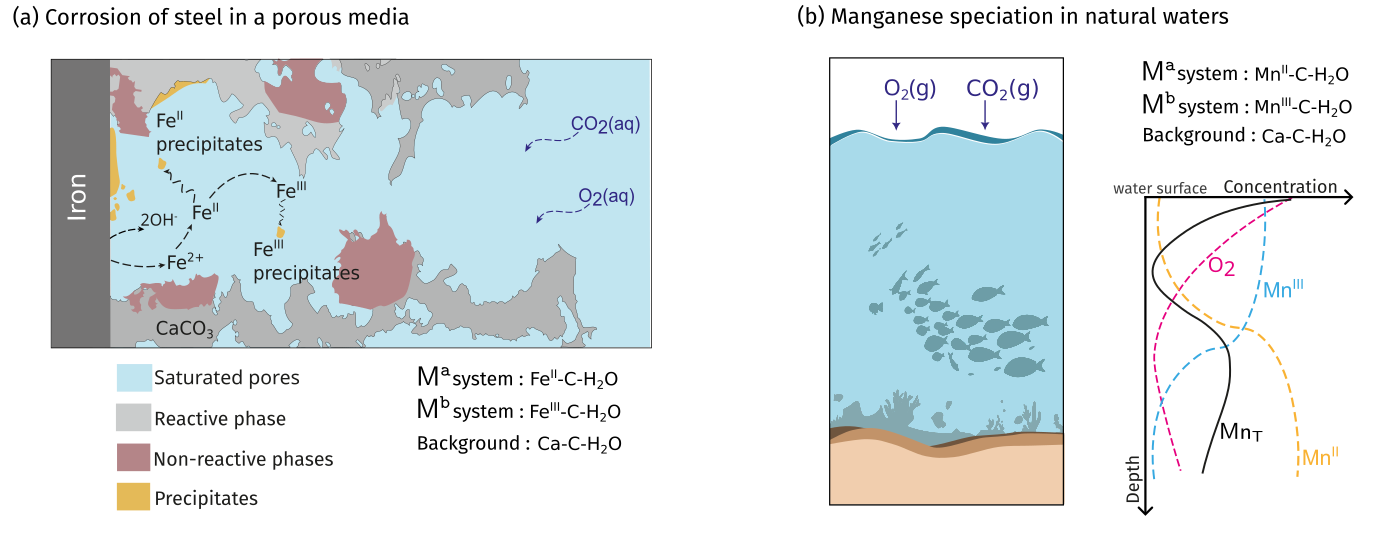}
    \caption{\textbf{Application examples of the proposed reactive transport framework:} (a) Case study I: On the fate of electrochemically dissolved iron in porous media. The interface between iron or steel and the porous medium is a multi-component system comprising of reactive phases, \ce{CaCO_3} (gray), non-reactive phases (brown), and aqueous phases such as pores saturated with moisture (blue). (b) Case study II: On the speciation of \ce{Mn} in natural waters. \ce{Mn} can exist in multiple oxidation states (\ce{Mn^{II}}, \ce{Mn^{III}}, \ce{Mn^{IV}}) and each of these species can play an important role in the biogeochemical cycles of several elements.}
    \label{fig:applications}
\end{figure}

We showcase the applicability and capabilities of our reactive transport framework through two examples: Case I: the fate of electrochemically dissolved iron in porous media (\Cref{fig:applications}a), and, Case II: manganese speciation in natural waters (\Cref{fig:applications}b). Iron (Fe) and manganese (Mn) are among the most abundant transition metals on earth and are critical elements in various socio-economically important man-made and natural aqueous environments. Their redox chemistry not only influences several biogeochemical cycles but also plays an important role in the functioning of critical infrastructure systems. In the selected examples, both Mn and Fe can exist in multiple oxidation states over different length and time scales and influence a multitude of local environmental conditions and processes. Therefore, knowledge of the spatial and temporal distribution of different species is critical. 

\subsection{Case I: Electrochemically dissolved iron in reactive porous media}\label{sec:corrosion}

Corrosion of steel or iron in porous media such as soils \cite{Neff2005CorrosionSystem,Cole2012TheSoils,Gardiner2002CorrosionMedia} or concrete ~\cite{Stefanoni2019KineticsMedia} proceeds by the anodic dissolution of Fe, releasing \ce{Fe^{2+}}, and the simultaneous cathodic reduction of water or oxygen (depending on \ce{pO_2}), releasing \ce{OH^{-}}. \ce{Fe^{2+}} can hydrolyse and form complexes with other anions, oxidise to \ferric{} species, precipitate as a solid \ferrous{} phase, and diffuse away from the steel surface (\Cref{fig:applications}a). Simultaneously, \ferric{} ions could precipitate as \ferric{} solid phases, sorb onto solid surfaces, and also diffuse away from the metal surface (\Cref{fig:applications}a). These processes are highly interdependent and are in kinetic competition with each other. Their individual rates are determined by the prevailing chemical environment (pH, concentration of other ionic species, \ce{pO_2}) and are governed by the thermodynamic stability of the aqueous and solid phases. Importantly, each of these processes can dynamically alter the chemistry of the local environment (\ref{sec:SM5-corrosion}), which thereby influences the rates of corrosion, oxidation, precipitation, sorption as well as shifts the thermodynamic equilibria. 

In solving such a complex system, current reactive transport models \cite{Stefanoni2018TheConcrete, Leupin2021AnaerobicInterface, Cole2012TheSoils} have two main limitations: a) they consider the rate of oxidation of \ferrous{} to \ferric{} to be constant and not evolving as a function of the prevailing chemistry of the aqueous phase (despite the pronounced influence of eg. pH \cite{Mundra2023AerobicSolutions}), and b) the reaction term cannot account for the thermodynamics of \ferrous{} and \ferric{} aqueous and solid phases simultaneously. In our framework, considering \ferrous{} oxidation reactions within the transient processes allows us to decouple the Gibbs free energy minimisation for metastable \ferrous{} and the more stable \ferric{} species. Ensuring the composition of the background species (in a porous medium such as soil or concrete) within each of the minimisation schemes is the same for \ferrous{} and \ferric{} species, we accurately predict the activity of the \ferrous{} and \ferric{} species at every timestep. This allows us to consider the kinetic rate constant for \ferrous{} oxidation as a function of the activity and speciation of \ferrous{}, and other parameters such as pH and \ce{pO_2}.

To illustrate this, we here employ the example of iron or steel corrosion in porous media~\cite{Stefanoni2019KineticsMedia,Neff2005CorrosionSystem,Cole2012TheSoils,Gardiner2002CorrosionMedia}, a longstanding challenge with a large socio-economic impact \cite{angst_challenges_2018}. For instance, consider corrosion in a porous medium such as steel in carbonated concrete fully saturated with an aqueous electrolyte (\Cref{fig:applications}a and \ref{sec:SM5-corrosion}) or corrosion of an archaeological artefact or steel pipes in soil (with the reactive soil particles here simplified as \ce{CaCO_3}). We apply our proposed framework to understand the influence of three key parameters, (a) porosity of the porous medium, (b) the concentration of reactive solid phases such as \ce{CaCO_3} surrounding the pores and (c) the presence or absence of oxygen, on the chemical evolution of the metal-porous medium interface as well as of the surrounding matrix.

\newpage
\thispagestyle{empty}

\begin{figure}[H]
    \centering
    \includegraphics[width=0.95\textwidth]{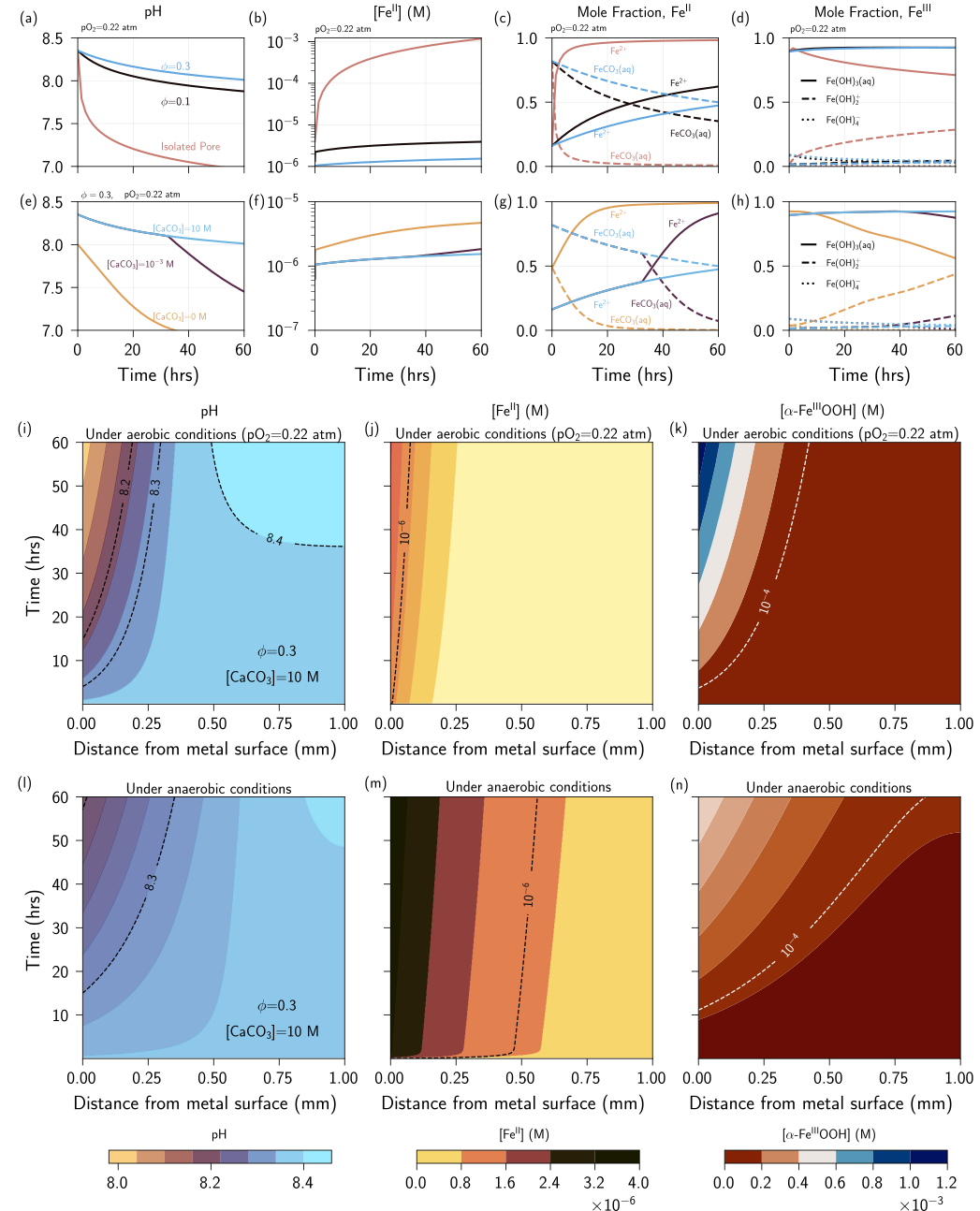}
    \caption{\textbf{Simulations of electrochemically dissolved iron in reactive porous media, revealing the temporal and spatial changes in pH, [Fe] as well as Fe speciation and precipitation as a function of the porosity, \ce{[CaCO_3]} and aeration condition:} (a-d) Modelled evolution of interfacial pH, [\ce{Fe^{II}}], speciation of \ce{Fe^{II}}, and speciation of \ce{Fe^{III}}, respectively, over time for three different average porosities of the porous medium (0.1, 0.3 and "isolated pore" - see \Cref{sec:method-iron}) under aerobic conditions. (e-h) Modelled interfacial pH, [\ce{Fe^{II}}], speciation of \ce{Fe^{II}}, and speciation of \ce{Fe^{III}} as a function of time and the amount of \ce{CaCO_3} (10 M, 0.001 M and 0 M) surrounding the pore network under aerobic conditions. Here, the average porosity is constant and equivalent to 0.3. (i-k) and (l-n) Spatial and temporal evolution of pH, [\ce{Fe^{II}}] and the solid iron (hydr)oxide precipitated in aerobic and anaerobic conditions, respectively. For (i-n), both the average porosity and the amount of \ce{CaCO_3} surrounding the pore network are fixed to 0.3 and 10 M, respectively. As the [\ferric{}] is governed by the solubility of goethite, the [\ferric{}] is in the order of \ce{10^{-13}} mol/L (see \Cref{fig:ferric-interface-conc-speciation} and \Cref{fig:ferric-conc-space}).}
    \label{fig:Influence of diffusivity}
\end{figure}

\newpage

Our simulations (\Cref{sec:method-iron}) suggest that the pH at the metal-porous medium interface is not constant and evolves over time depending on the porosity (\Cref{fig:Influence of diffusivity}a), the amount of reactive solid phases such as \ce{CaCO_3} surrounding the pores (\Cref{fig:Influence of diffusivity}e), and the presence (\Cref{fig:Influence of diffusivity}i) or absence (\Cref{fig:Influence of diffusivity}l) of dissolved oxygen~(\ref{sec:SM5-corrosion}). The interfacial pH decreases with time (\Cref{fig:Influence of diffusivity}a, e, i, l), which can be attributed to various pH dependent processes such as oxidation of \ferrous{} ions ~\cite{Mundra2023AerobicSolutions, Stumm1961OxygenationIron, Pham2008OxygenationPathways, Millero1987OxidationStrength, Weiss1935ElektronenubergangsprozesseLosungen}, hydrolysis and speciation of \ferrous{} and \ferric{} ions ~\cite{furcas_solubility_2022}, and the precipitation of thermodynamically stable \ferrous{} or \ferric{} solid phases ~\cite{furcas_solubility_2022, Furcas2023TransformationPH} (see \ref{sec:SM5-corrosion}). As the interfacial pH decreases, the solubility of \ferrous{} increases~\cite{furcas_solubility_2022} and therefore, the concentration of \ferrous{} species at the interface increases (\Cref{fig:Influence of diffusivity}b, f, j, m). However, as the concentration of \ferrous{} is also dynamically controlled by the pH and \ce{pO_2} dependent oxidation rate of \ferrous{} (or source of \ferric{}), the concentration of \ferrous{} does not necessarily follow the solubility limit of \ferrous{} w.r.t. \ce{Fe(OH)_2}(s) or \ce{FeCO_3} (s) (\ref{sec:si-ferrous-stability}), which is a feature traditional modelling frameworks struggle accounting for. We observe this when comparing \Cref{fig:Influence of diffusivity}j and \Cref{fig:Influence of diffusivity}m, where the [\ferrous{}] at the metal-porous medium interface is much higher in anaerobic environments than aerobic environments, even though the pH is lower in the case of the latter. Furthermore, our framework allows untangling the aqueous speciation of \ferrous{} and \ferric{} temporally and spatially, and assess the concentration of each of the species participating in the oxidation reactions. The majority of the interfacial \ferrous{} initially exists predominantly in the form of \ce{FeCO_{3}}(aq) and \ce{Fe^{2+}} (\Cref{fig:Influence of diffusivity}c and \Cref{fig:Influence of diffusivity}g). Hydrolysis products of \ce{Fe^{2+}} such as \ce{FeOH^{+}}, \ce{Fe(OH)_{2}}(aq), and \ce{Fe(OH)_{3}^{-}} are only present in trace quantities at the pH of the interface. Upon time and a decrease in the pH, non-hydrolysed \ce{Fe^{2+}} becomes the predominant aqueous \ferrous{} species, in line with the thermodynamic stability of different aqueous \ferrous{} species. Similarly, \ferric{} initially exists as \ce{Fe(OH)_{3}}(aq), \ce{Fe(OH)_{4}^{-}} and \ce{Fe(OH)_{2}^{+}} (\Cref{fig:Influence of diffusivity}d, h). Upon a decrease in the pH, the concentrations of \ce{Fe(OH)_{4}^{-}} and \ce{Fe(OH)_{3}} (aq) reduce, whereas the concentration of \ce{Fe(OH)_{2}^{+}} increases (\Cref{fig:Influence of diffusivity}d, h). Our framework allows to track spatial and temporal changes in the speciation of \ferrous{} and \ferric{} and the activity of each of the aqueous species as the pH and aqueous chemistry evolve.

While the interfacial chemistry is important, the processes occurring further away from the metal surface are equally relevant for assessing the long-term corrosion behaviour of the system~\cite{Pundir2023AnMedia, Stefanoni2018TheConcrete}. Owing to the slower rate of \ferrous{} oxidation to \ferric{} under anaerobic environments \cite{Mundra2023AerobicSolutions}, model results demonstrate higher [\ferrous{}] at any point in space and time along the distance of 1~mm away from the metal surface under anaerobic conditions (\Cref{fig:Influence of diffusivity}j) than aerobic conditions (\Cref{fig:Influence of diffusivity}m). As a result (and in agreement with experimental observations \cite{Stefanoni2019KineticsMedia}), the precipitation of \ferric{} (hydr)oxide (from aqueous \ferric{}) $-$ goethite $-$ is predicted to occur further away from the metal-porous medium interface in anaerobic environments, albeit at lower quantities when compared to the case of aerobic conditions (\Cref{fig:Influence of diffusivity}k, n). Consequently, the reduction in the pH of the aqueous phase over time (lower than the initial pH of 8.4) within the porous medium is restricted to $\approx$ 300 \ce{\mu}m within the first 60 hours in the case of aerobic conditions (\Cref{fig:Influence of diffusivity}i). Conversely, the pH of the aqueous phase in anaerobic conditions is lowered all along the distance of 1 mm away from the metal-porous medium interface (\Cref{fig:Influence of diffusivity}l). As we assume \ferric{} solid phases to precipitate instantaneously, the concentration of \ferric{} (\Cref{fig:ferric-conc-space}) is dictated by the pH dependent solubility of the precipitating solid phase $-$ goethite.

Given the empirical nature of existing models, such as those used for predicting the corrosion of steel in soil or cementitious media \cite{Leupin2021AnaerobicInterface, Cole2012TheSoils,Gardiner2002CorrosionMedia}, we view the proposed model framework as a potential foundation for the development of more fundamental and mechanistic models.

\subsection{Case II: Speciation of manganese in natural waters}\label{sec:app-mn}\label{sec:manganese}

Manganese (\ce{Mn}) plays an important role in natural coastal aqueous ecosystems, where it serves as a mediator in several redox cycles of elements such as iron, oxygen, sulfur, carbon, nitrogen, among others \cite{Burdige1993TheSediments, Landing1987TheOcean}. Though \ce{Mn^{IV}} (in the form of a solid oxide phase) is the thermodynamically most favoured state in near-neutral aqueous environments, \ce{Mn^{II}} and in particular, \ce{Mn^{III}} constitute a large fraction of the soluble Mn-pool \cite{Luther2018ReductionSteps, Fritsch1996ThermodynamicOxides}. The oxidation of \ce{Mn^{II}} to \ce{Mn^{IV}} and the reduction of \ce{Mn^{IV}} to \ce{Mn^{II}} involves the formation of a \ce{Mn^{III}} intermediate that could either disproportionate or readily complexes with ligands (depending on the pH) \cite{Oldham2020TheOcean, Luther2018ReductionSteps, Trouwborst2006SolubleZones, Morgan2005KineticsSolutions, Langen1996OxidationConcentrations}. \ce{Mn^{III}} is an important redox species that can act as a oxidant or reductant, and is responsible for the existence of suboxic zones in estuaries, oceans and sedimentary porewaters of lakes \cite{Trouwborst2006SolubleZones}. The reactions surrounding the redox cycling of Mn in marine waters comprise several oxidative and reductive pathways (photochemical, biological, inorganic, \ce{pO_2}, surface-catalysed, pH, among others) each affecting the reactivity of Mn and other biogeochemical cycles \cite{Grill1982KineticWaters, Richard2013KineticsSediments, VanCappellen1996CyclingManganese}. In this context, we apply our reactive transport framework to study the speciation of Mn along a 30-meter-deep water column (\Cref{fig:applications}b), supersaturated in \ce{CaCO_3}. We investigate scenarios of elevated levels of \ce{pCO_2} in the atmosphere, potentially originating from multiple sources such as increased fossil fuel combustion, leakage of captured \ce{CO_2}, or microbial emissions. 

We consider an initial distribution of \ce{Mn^{II}} along the depth, similar to those profiled in the literature~\cite{Johnson1996OnMinimum, Johnson1992ManganeseMinimum} (\Cref{fig:supp-mn}b). At the surface of the water column, we suppose an inward flux of \ce{Mn^{II}} $-$ equivalent to $10^{-11}$ mol/s/m$^2$~\cite{Johnson1992ManganeseMinimum, Morgan2005KineticsSolutions}, that mimics natural processes such as lithogenic dust deposition. Here, we assume the activity of dissolved oxygen to vary spatially and decreases with depth (as reported in~\cite{Grengs2024DirectEnvironments}, assumed constant throughout the simulation period) (\Cref{fig:supp-mn}c). We consider three concentrations of \ce{CO_2}, each corresponding to a partial pressure of $10^{-3.5}$ atm, $10^{-2.5}$ atm and $10^{-1.5}$ atm. The entirety of the water column is assumed to be in equilibrium with \ce{pCO_2}. To observe the long-term implication of elevated \ce{CO_2} levels, we simulate a time period of 340 days. We use commonly reported oxidation and reduction kinetic constants \cite{Morgan2005KineticsSolutions, Klewicki1996TheComplexes} as a function of pH and \ce{pO_2} (\Cref{fig:supp-mn}a) to consider oxidation of \ce{Mn^{II}} and reduction of \ce{Mn^{III}}, respectively, along with other transient processes such as diffusion and advection. This enables us to simultaneously solve for speciation of both \ce{Mn^{II}} and \ce{Mn^{III}}, as a function of depth and time. Given the lack of thermodynamic data and redox rate laws concerning aqueous \ce{Mn^{IV}} species, we currently cannot predict its speciation, activity or the amount of solid \ce{Mn^{IV}} phases formed. Thus, \ce{Mn^{IV}} is excluded in this example. 

\begin{figure}[H]
    \centering
    \includegraphics[width=0.95\textwidth]{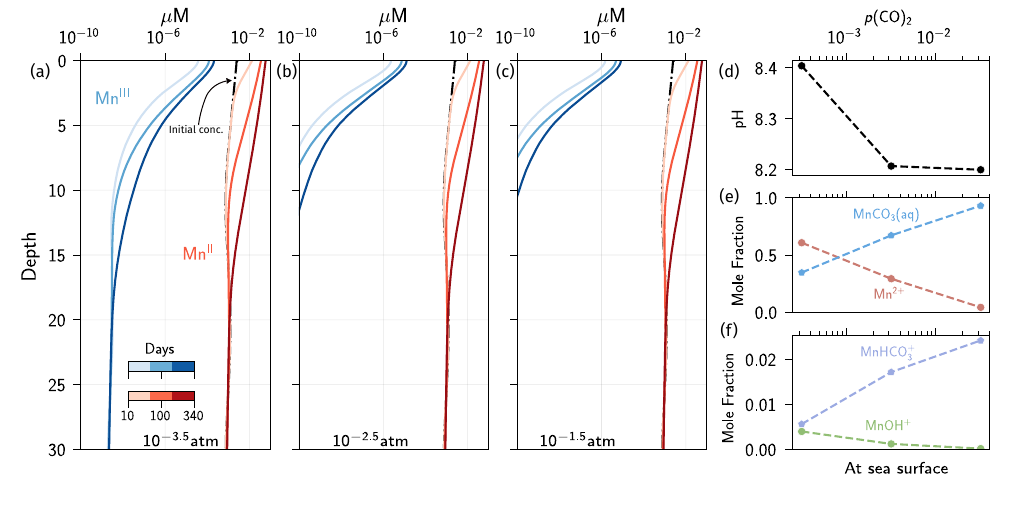}
    \caption{\textbf{Influence of elevated levels of \ce{CO_2} concentration on Mn speciation in natural waters: } Distribution of \ce{Mn^{II}} (red) and \ce{Mn^{III}} (blue) along the depth of the water column for three different \ce{pCO_2} (a) $10^{-3.5}$ atm, (b) $10^{-2.5}$ atm and (c) $10^{-1.5}$ atm at 10 days, 100 days and 340 days. Effect of \ce{pCO_2} on (d) pH and (e, f) the mole fractions of relevant \ce{Mn^{II}} complexes, after a period of 340 days and for the spatial location at the surface of the sea. The black dashed line represents the initial ($t$ = 0) concentration of \ce{Mn^{II}} as a function of depth (\Cref{fig:supp-mn}b). A 0.1 M concentration of \ce{CaCO_3} is assumed for all cases.}
    \label{fig:oceanic-waters}
\end{figure}

The results show that as \ce{Mn^{II}} enters the water column, the higher \ce{pO_2} at shallow depths facilitates the faster oxidation of \ce{Mn^{II}} to \ce{Mn^{III}} (\Cref{fig:oceanic-waters}a, b, c). Whereas at lower depths, the slower oxidation of \ce{Mn^{II}} and rapid reduction of \ce{Mn^{III}}, limits the activity of \ce{Mn^{III}} and only causes negligible changes in the initial distribution of \ce{Mn^{II}}. Over time, the diffusion of \ce{Mn^{II}} and \ce{Mn^{III}} increases their respective activities along the depth of the water column. Similar to the \ferrous{}-\ferric{} system discussed in \Cref{sec:corrosion}, our reactive transport framework allows us to map the influence of hydrolysis, speciation and precipitation of \ce{Mn^{II}} and \ce{Mn^{III}} on the pH across the entire spatial domain. Over the course of 340 days, the concentration of both \ce{Mn^{II}} and \ce{Mn^{III}} always remains below the solubility of solid \ce{Mn^{II}} and \ce{Mn^{III}} phases. Therefore, our model does not predict the precipitation of solid phases. Additionally, the high acid-neutralising capacity of \ce{CaCO_3} (s) considered here does not allow for the pH to change across the time and spatial domains we investigated. 

From our simulations we infer that an increase in the \ce{pCO_2} in the atmosphere results in a decrease in the pH of the aqueous system (\Cref{fig:oceanic-waters}d). While the reduction in the pH does not seem to influence the [\ce{Mn^{II}}] along the depth for the three \ce{pCO_2} studied here (\Cref{fig:oceanic-waters}a, b, c), speciation of \ce{Mn^{II}} changes significantly. \ce{Mn^{2+}} and \ce{MnCO_3} (aq) constitute for roughly 60 $\%$ and 40 $\%$ of the total \ce{Mn^{II}}, respectively, when the pH is $~$ 8.4 (when \ce{pCO_2} = $10^{-3.5}$ atm) (\Cref{fig:oceanic-waters}e). Whereas, at higher \ce{pCO_2}, the solubility of aqueous \ce{Mn^{II}} is controlled by the solubility of \ce{MnCO_3} (aq), and \ce{Mn^2+} only exists in minor quantities (\Cref{fig:oceanic-waters}e). \ce{MnHCO_3^+} and \ce{MnOH^+} only exist in trace quantities (\Cref{fig:oceanic-waters}f), and do not contribute significantly to the overall solubility of \ce{Mn^{II}} in the pH range of 8.2 - 8.4. Finally, our model indicates (\Cref{fig:oceanic-waters}a-c) that elevated \ce{pCO_2} can decrease the amount of mobile \ce{Mn^{III}} in natural waters with similar composition as assumed here. At a lower pH, the decreased rate of \ce{Mn^{II}} oxidation in conjunction with a higher rate of \ce{Mn^{III}} stabilises \ce{Mn^{II}} species in solution. 

\section{Outlook}\label{sec-outlook}
We set out the fundamental basis for a methodology to gain insights into the evolution of the aqueous chemistry of man-made and natural environments. This opens up several avenues for future research. The innate flexibility within our framework allows future work to integrate key transient processes such as the sorption of aqueous species onto solid surfaces, transformations of solid phases from thermodynamically less stable to more stable forms, and the influence of precipitation/dissolution on the pore structure, without compromising on the thermodynamic equilibrium calculations of the aqueous and solid phases of metals in several oxidation states. Moreover, we consider it a major advantage that our framework can also account for the precipitation of solid phases containing metals in mixed oxidation states, for example, \ferrous{} and \ferric{} in \ce{Fe_3O_4}(s) or \ce{Mn^{II}}, \ce{Mn^{III}} and \ce{Mn^{IV}} in \ce{MnO_x}(s)/\ce{Mn_3O_4}(s). We propose to take into account the precipitation of such solid phases along with other transient processes, provided availability of accurate experimentally determined rate laws. As with most reactive transport models, predicting long-term behaviour is computationally intensive, particularly when large time steps are required. This presents a challenge that could be addressed by developing more efficient and innovative approaches to speed up thermodynamic calculations \cite{Leal2020AcceleratingCalculations, guan_differentiable_2022}.

Our reactive transport framework showcases a novel approach in addressing the challenges in mapping the spatial and temporal distribution of elements across multiple redox states in aqueous systems. This capability is particularly valuable in fields such as toxicity assessment, groundwater remediation, and corrosion, among others, where metals can exist in multiple oxidation states and significantly influence aqueous chemistry and transport behaviour.

\section{Methods}\label{sec:method}

\subsection{Operator-split algorithm for thermodynamic modelling}
As experimental data are scarce on individual formation rates $\kinetic_i$ of different $\metal^{\bstate}$ complexes from $\metal^{\astate}$ complexes, it is not possible to model the chemical reactions as a whole, that is, under partial equilibrium conditions ~\cite{leal_overview_2017}. To overcome this problem, we propose using an operator-split approach to simulate chemical reactions in two separate but parallel chemical systems (see \Cref{fig:methodology}). In the first system, we consider the speciation of $\metal^{\astate}$ and its precipitation, and in the second system, we consider the speciation and precipitation of $\metal^{\bstate}$. We consider the oxidation of $\metal^{\astate}$ to $\metal^{\bstate}$ by introducing an auxiliary variable [$\metal^{\bstate}$], which represents the sum of all species in the oxidation state $\bstate$. Similarly, to account for the reduction of $\metal^{\bstate}$ to $\metal^{\astate}$, we consider an auxiliary variable [$\metal^{\astate}$], which represents the sum of all species in the oxidation state $\astate$. The formation of each of the auxiliary variables is kinetically controlled with an overall rate constant $\kinetic$ that is representative of the combined formation rates of each $\metal^{\bstate}$ complex \ie{} $\kinetic = f(\kinetic_1, \kinetic_2, ... \kinetic_{n})$. Our choice is motivated by the fact that in experiments, the parameter $\kinetic$ is measurable instead of the individual parameters $\kinetic_i$~\cite{Mundra2023AerobicSolutions}. We use an iterative approach to solve the two systems. Initially, the focus is on the first chemical system, which involves the equilibrium speciation and precipitation of $\metal^{\astate}$. We employ the Gibbs minimisation principle for this purpose \cite{leal_computational_2016}. The pH value obtained from this process, along with the equilibrium amount of background species (excluding $\metal$ species), serves as the initial condition for the second chemical system. The second system is then solved to determine the speciation and precipitation of $\metal^{\bstate}$. At this point, the pH value obtained from the second system is checked. If the pH values of the first and second systems are within a specified tolerance, it is assumed that the two systems have reached equilibrium. However, if the pH value from the second system is outside the tolerance, it is passed along with the amount of background species to the first system for another iteration. This process is repeated until the pH values of both systems are within the chosen tolerance. \Cref{fig:schematic-thermo} shows a schematic representation of the operator-split algorithm for the thermodynamic modelling. It is worth noting that the proposed operator-split approach is robust, and the order in which the systems are solved does not impact the results.

\subsection{Using Gibbs minimisation for solving a chemical system}
For a given chemical system, the Gibbs minimization problem is formulated as:
\begin{equation}\label{equation-1}
    \underset{n}{\mathrm{min}}~G = n_i\mu_i \quad \mathrm{subject~to} \begin{cases}
        A_{ij}n_j =b_i \\
        n_j \geq 0 \\
        \ce{CO2} = p~\mathrm{atm}
    \end{cases}     
 \end{equation}
where $n_i$ is the moles of species, $\mu_i$ is the chemical potential of $i^{\mathrm{th}}$ specie,  $A_{ij}$ is the stoichiometric coefficient of $i^{\mathrm{th}}$ species in $j^{\mathrm{th}}$  reaction and $b_j$ is the moles of $j^{\mathrm{th}}$ element. The chemical potential function of the $i^{\mathrm{th}}$ species is given as $ \mu_i(n) = \mu_i^{\circ} + \mathrm{RT} \ln{a_i(n)} $ where $\mu_i^\circ$ is the standard chemical potential function and $a_i$ is the activity function of the $i^{\mathrm{th}}$ species. The activity of the specie $i^{\mathrm{th}}$ is defined as: $\ln{a_i} = \ln{\psi_i} + \ln{x_i} + \ln{c}$ where for aqueous species $\psi_i$ is equal to the activity coefficient $\gamma_i$ and for gaseous specie $\psi_i$ represents the fugacity coefficient. Similarly, $c$ is a constant with a value of 1 for aqueous species and a value of $c=P/P^\circ$ for gaseous species. We employ the ideal activity model to calculate the coefficients of aqueous activity and gaseous activity~\cite{leal_overview_2017, leal_computational_2016}. The oxidation and reduction kinetic rate constants are dependent on the solution's $\pH$ and the oxygen state (aerobic or anaerobic) of the system. 
We use Reaktoro~\cite{leal_computational_2016} to solve the constrained minimisation problem.

\subsection{Coupling diffusion process with chemical reactions}
We couple the proposed operator-split approach with a multi-species diffusion process to simulate and understand the complex mechanisms at the metal surface. We employ a sequential non-iterative approach~\cite{Sun2019, damiani_framework_2020} to couple the chemical and diffusion processes. The coupled diffusion equations for $\metal^{\astate}$ aqueous complexes, $\metal^{\bstate}$ aqueous complexes and auxiliary variables, [$\metal^{\astate}$] and [$\metal^{\bstate}$], read:
\begin{align}\label{eq:diffusion-3}
 \sum_i \dt{\metal}^{\astate}_i (\vector{x}, t) &=  \sum_i \nabla D_{\metal^{\astate}_{i}} \cdot{} \nabla \metal^{\astate}_{i}(\vector{x}, t)  - \ox \sum_i \metal^{\astate}_i (\vector{x}, t)~, \\
  \sum_j \dt{\metal}^{\bstate}_j (\vector{x}, t) &=  \sum_j \nabla D_{\metal^{\bstate}_{j}} \cdot{} \nabla \metal^{\bstate}_{j}(\vector{x}, t)  - \red \sum_j \metal^{\bstate}_j (\vector{x}, t)~, \\
  \dt{\metal}^{\bstate} (\vector{x}, t) &= \nabla D_{\metal^{\bstate}} \nabla \metal^{\bstate} (\vector{x}, t) + {\ox\sum_{i=1}^{m} \metal^\astate_{i}} - {\red \metal^{\bstate}}~, \\
    \dt{\metal}^{\astate} (\vector{x}, t) &= \nabla D_{\metal^{\astate}} \nabla \metal^{\astate} (\vector{x}, t) - {\ox \metal^{\astate}} + {\red\sum_{j=1}^{n}  \metal^\bstate_{j}}~,
\end{align}
For the other aqueous species (\ie{} background species in our framework), we employ the above equation for the diffusion process but without any source or sink term. We perform a sensitivity analysis to ensure that the time-step size $\Delta t$ does not affect the reactive transport modelling; see \Cref{fig:time-step}. Based on the sensitivity analysis, we chose a time step of 5 s for the simulations in the paper. The concentration of each species (both aqueous and solid) is expressed as moles per litre. We compute the total number of moles of each species in a finite element and divide it by the total volume of that spatial finite element.

\subsection{Iron in porous media}\label{sec:method-iron}

We perform reactive transport simulations over a length of 1 mm along the steel-concrete interface. We assume that the iron corrodes at a constant rate of $1 \mu\ce{A/cm^2}$ leading to a constant flux of \ce{Fe^{2+}} ions from the steel interface. The rate at which \ce{Fe^{II}} oxidises to \ce{Fe^{III}} depends on the pH, \ce{pO_2} according to the values reported in \cite{Mundra2023AerobicSolutions}. We approximate the complex pore structure of a fully carbonated concrete and assume three microstructural scenarios at the steel-concrete interface (\Cref{fig:supp-fe}a), with different average porosities. In the first two scenarios, we consider two average porosities, namely 0.1 and 0.3, perpendicular to the steel (\Cref{fig:supp-fe}a). The third scenario considers the case of a significantly small pore at the steel-concrete interface completely filled with the pore solution of carbonated concrete. In other words, this represents a thin pore along the length of the steel reinforcement, where the diffusion of any aqueous species is not considered (we assume $D$ to be 0 in Equations 3 $-$ 5), and we term this an “isolated pore” (or could be considered a “crevice” in the field of corrosion science) (\Cref{fig:supp-fe}a). The solids surrounding the pores are solely constituted of a reactive species - \ce{CaCO_3} (s) - in equilibrium with the aqueous phase and can congruently dissolve in case of changes in the pH of the electrolyte. We vary the initial amount of \ce{CaCO_3} (s) present in our simulations from 10 mol/L to 0.001 mol/L to 0 mol/L to study the influence of the acid-neutralising capacity of \ce{CaCO_3} surrounding the pore network. While assessing the influence of the acid-neutralising capacity of \ce{CaCO_3}, we fix the porosity to a value of 0.3. As the concentration of dissolved oxygen can influence the rate at which \ce{Fe^{II}} oxidises, we consider the concrete cover to be either aerobic and anaerobic. The influence of oxygen on the spatial and temporal evolution of the chemistry of the concrete cover is assessed using experimentally determined pseudo-first order kinetic constants of \ce{Fe^{II}} oxidation in aerobic and anaerobic environments \cite{Mundra2023AerobicSolutions} (\Cref{fig:supp-fe}c). While assessing the influence of oxygen, we fix the amount of \ce{CaCO_3} surrounding the pores to 10 M, and the average porosity to 0.3.

\subsection{Manganese in natural waters}\label{sec:method-Mn} 

We consider a sea column 30 m in depth. We assume that the partial pressure of $\ce{O_2}$ decreases exponentially as  a function of depth from the sea surface given as $\ce{pO_2}(y) = 0.22 \exp{-y/5}$ where 0.22 is the atmospheric pressure of \ce{O_2} and $y$ is the depth from the sea surface \Cref{fig:supp-mn}c.  We assume that the partial pressure of $\ce{CO_2}$ does not change along the depth of the sea column. In \Cref{sec:app-mn}, we use 3 different values for $\ce{pCO_2}$ and assume that the entire sea column is in equilibrium with elevated $\ce{pCO_2}$ and thus the entire column has the same $\ce{pCO_2}$. The overall pseudo first-order oxidation and reduction kinetic rate constants are determined from the experimentally recorded values from the literature~\cite{Morgan2005KineticsSolutions, Klewicki1996TheComplexes} and is a function of the pH and $\ce{pO_2}$.

\section{Acknowledgements}
The authors would like to thank the European Research Council (ERC) for the financial support provided under the European Union’s Horizon 2020 research and innovation programme (ERC Starting Grant agreement: Taming corrosion, 848794). The authors would also like to thank the Swiss National Science Foundation (SNSF) for providing financial support for the work under grants (PP00P2\_194812); and Dr Fabio E. Furcas and Luca Michel for helpful discussions. 

\section{Author Contributions}
S.M - conceptualisation, analysis, investigation, methodology, writing - original draft, writing - review and editing. M.P - conceptualisation, analysis, investigation, methodology, software, validation, writing - original draft, writing - review and editing. B.L - analysis, writing - review, D.S.K and U.M.A - conceptualisation, writing - review, funding acquisition, supervision. 

\section{Ethics Declaration}
The authors declare no competing interests. 

\newpage

\newpage

\iftrue
\clearpage
\renewcommand{\appendixname}{Supplementary Information}
\appendix
\renewcommand{\thesection}{Supplementary Note \arabic{section}}    %
\renewcommand{\thefigure}{SI~\arabic{figure}}
\setcounter{figure}{0}
\renewcommand{\thetable}{SI~\arabic{table}}
\setcounter{table}{0}

\newpage

\section{Methods}\label{sec:methodsSM-1}

\begin{figure}[H]
    \centering
    \includegraphics[width=0.9\textwidth]{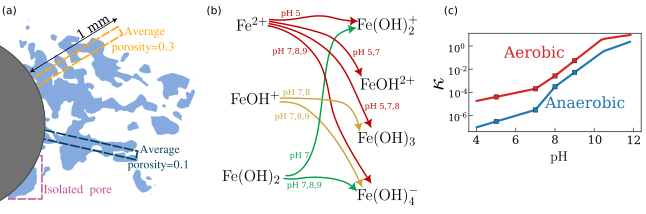}
    \caption{(a) Schematic representation of the steel concrete interface, with three average porosities. (b) Oxidation schematic with different \ce{Fe^{II}} and \ce{Fe^{III}} species~\cite{Mundra2023AerobicSolutions}. (c) Pseudo-first order oxidation kinetics (in \ce{sec^{-1}}) of \ce{Fe^{II}} as a function of pH (aerobic and anaerobic)~\cite{Mundra2023AerobicSolutions}.}
    \label{fig:supp-fe}
\end{figure} 

\begin{figure}[H]
    \centering
    \includegraphics[width=0.9\textwidth]{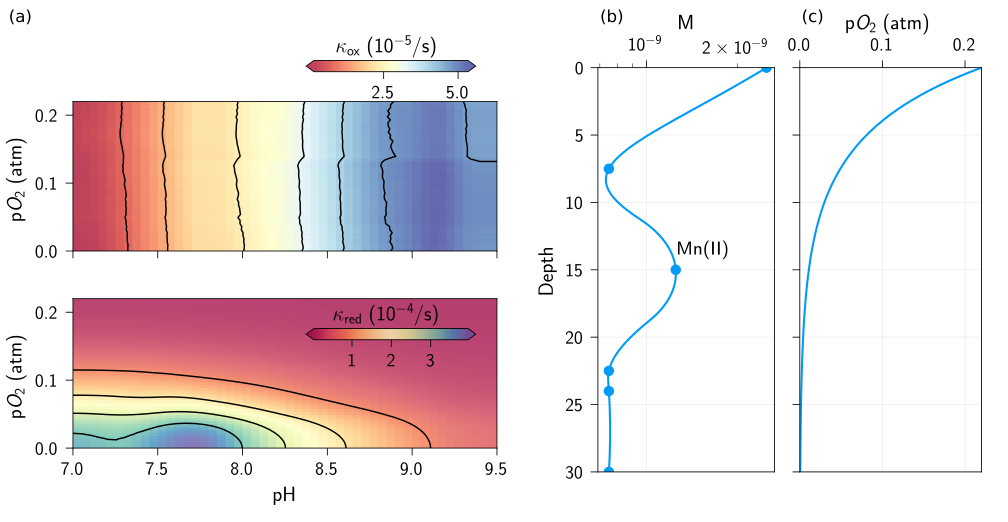}
    \caption{(a) \ce{Mn^{II}} oxidation and \ce{Mn^{III}} reduction kinetics interpolation vs. pH and \ce{pO_2} (data from~\cite{Morgan2005KineticsSolutions, Klewicki1996TheComplexes}). (b) Initial conditions of \ce{Mn^{II}} distribution in oceans~\cite{Johnson1996OnMinimum}. (c) Oxygen concentration vs. depth~\cite{Johnson1996OnMinimum}.}
    \label{fig:supp-mn}
\end{figure}

\newpage

\section{Thermodynamic database}\label{sec:SM2-thermodatabase}

Thermodynamic data for the solid, aqueous and gaseous species considered in this study for the Fe-C-\ce{H_2O} system.

\begin{table}[H]
\begin{adjustbox}{width=1\textwidth}
\begin{tabular}{llllllll}
  \hline
  \textbf{Species} & \textbf{MW} (g/mol)  & $\mathbf{\Delta_fG^\circ_m}$, $\si{\kilo\joule\per\mole}$ & $\mathbf{\Delta_fH^\circ_m}$, $\si{\kilo\joule\per\mole}$  & $\mathbf{S^\circ_m}$, $\si{\joule\per\mole\per\kelvin}$  & $\mathbf{C_{p,m}^\circ}$, $\si{\joule\per\mole\per\kelvin}$  & $\mathbf{V^\circ}$, $\si{\joule\per\bar}$ & \textbf{Ref.} \\
  \hline
  $\alpha$ - \ce{Fe0} & $55.8450$ & $0$ & $0$ & $27.085 \pm 0.160$ & $25.084 \pm 0.500$ & $0.7092 \pm 0.0004$ & \cite{Lemire2013Chemical1, Robie1995ThermodynamicTemperatures}\\
  \ce{Fe^2+} & $55.8450$ & $-90.719 \pm 0.641$ & -90.295 $\pm$ 0.522 & $-102.171 \pm 2.770$ & $-23.000 \pm 10.00$ & $-2.264$ & \cite{Lemire2013Chemical1} \\
  \ce{Fe^3+} & $55.8450$ & $-16.226 \pm 0.650$ & -50.056 $\pm$ 0.973 & $-282.404 \pm 3.927$ & $-108.000 \pm 20.00$ & $-3.779$ & \cite{Lemire2013Chemical1} \\
  \\
  \textbf{Fe(II) hydrolysis products} \\
  \hline
  \ce{Fe(OH)^+} & $72.8523$ & $-273.996 \pm 0.859^\text{a}$ & -321.525 $\pm$ 0.104 & $-29.628 \pm 4.851$ & $63.061$ & $-1.671$ & \cite{Brown2016HydrolysisIons}\\
  \ce{Fe(OH)_2} (aq) & $89.8597$ & $-447.798 \pm 0.791^\text{a}$ & -546.555 $\pm$ 1.131 & $31.930 \pm 4.941$ & & & \cite{Brown2016HydrolysisIons}\\
  Fe(O) (aq) (+ \ce{H2O (l) = Fe(OH)2} (aq))  & $71.8444$ & $-210.694 \pm 0.792$ & -260.725 $\pm$ 1.131 & $-38.020 \pm 4.940$ & $-0.013$ & $-1.649$ & \cite{Brown2016HydrolysisIons,furcas_solubility_2022}\\
  \ce{Fe(OH)_3^-} & $106.867$ & $-615.492 \pm 1.077$ & -807.485 $\pm$ 2.264 & $-47.406 \pm 8.579$ & & & \cite{Brown2016HydrolysisIons}\\
  \ce{FeO_2H^-} (+ \ce{H2O} (l) = \ce{Fe(OH)_3^-)} & $88.8518$ & $-378.388 \pm 1.077$ & -521.655 $\pm$ 2.264 & $-117.356 \pm 8.579$ & $92.844$ & $-1.347$ & \cite{Hummel2023The2020, Brown2016HydrolysisIons,furcas_solubility_2022}\\
  \\
 \textbf{Fe(III) hydrolysis products} \\
  \hline
  \ce{FeOH^2+} & $72.8523$ & $-240.772 \pm 0.661^\text{a}$ & -292.586 $\pm$ 1.144 & $-109.344 \pm 5.503$ & $-33.693$ & $-2.534$ & \cite{Brown2016HydrolysisIons}\\
  \ce{Fe(OH)_2^+} & $89.8597$ & $-457.841 \pm 0.869^\text{a}$ & -550.200 $\pm$ 1.00 & $-11.848 \pm 4.891$ & & & \cite{Brown2016HydrolysisIons}\\
  \ce{FeO+} (+ \ce{H2O} (l) = \ce{Fe(OH)_2^+}) & $71.8444$ & $-220.737 \pm 0.870$ & -264.300 $\pm$ 1.00 & $-81.798 \pm 4.890$ & $-200.942$ & $-4.202$ & \cite{Hummel2023The2020, Brown2016HydrolysisIons,furcas_solubility_2022}\\
  \ce{Fe(OH)_3} (aq) & $106.867$ & $-657.557 \pm 1.625^\text{a}$ & -761.246 $\pm$ 4.899 & $183.423 \pm 17.61$ & & & \cite{Brown2016HydrolysisIons}\\
  \ce{FeO_2H} (aq) (+\ce{H2O} (l) = \ce{Fe(OH)3} (aq)) & $88.8518$ & $-420.453 \pm 1.625$ & $-475.416 \pm 4.899$ & $113.473 \pm 17.61$ & $-312.138$ & $-0.721$ & \cite{Brown2016HydrolysisIons,furcas_solubility_2022}\\
  \ce{Fe(OH)_4^-} & $123.874$ & $-841.348 \pm 1.474^\text{a}$ & -1046.58 $\pm$ 2.052 & $76.237 \pm 9.067$ & & & \cite{Brown2016HydrolysisIons}\\
  \ce{FeO_2^-} (+ \ce{2H2O} (l) = \ce{Fe(OH)_4^-}) & $87.8438$ & $-367.140 \pm 1.474$ & -474.916 $\pm$ 2.052 & $6.287 \pm 9.067$ & $-234.928$ & $0.045$ & \cite{Brown2016HydrolysisIons,furcas_solubility_2022}\\
  \ce{Fe_2(OH)_2^4+} & $145.705$ & $-490.050 \pm 1.362^\text{a}$ & -641.672 $\pm$ 9.698 & $-379.664 \pm 33.48$ & & & \cite{Brown2016HydrolysisIons}\\
  \ce{Fe_3(OH)_4^5+} & $235.564$ & $-961.133$ & $-1234.49$ & $-490.138$ & & & \cite{Hummel2023The2020}\\
  \\
  \textbf{Fe(II) chloride complexes} \\
  \hline
  \ce{FeCl^+} & $91.298$ & $-216.228 \pm 4.613^\text{a}$ & -235.825 $\pm$ 1.848 & $7.563 \pm 16.574$ & $86.492$ & $0.085$ & \cite{Lemire2020Chemical2}\\
  \ce{FeCl_4^2-} & $197.657$ & $-584.735 \pm 3.352^\text{a}$ & -687.215 $\pm$ 2.585 & $260.229 \pm 7.571$ & & & \cite{Lemire2020Chemical2}\\
  \\
  \textbf{Fe(III) chloride complexes} \\
  \hline
  \ce{FeCl^2+} & $91.298$ & $-156.119 \pm 0.873$ & -194.656 $\pm$ 4.703 & $-121.309 \pm 16.04$ & $14.828$ & $-2.286$ & \cite{Lemire2020Chemical2}\\
  \ce{FeCl_2^+} & $126.751$ & $-291.332 \pm 1.442^\text{a}$ & -342.386 $\pm$ 8.864 & $13.596 \pm 33.631$ & $300.718$ & $1.027$ & \cite{Lemire2020Chemical2}\\
  \ce{FeCl_3} (aq) & $162.204$ & $-415.870 \pm 1.681$ & -494.666 $\pm$ 11.34 & $97.436 \pm 50.866$ & $368.217$ & $3.594$ & \cite{Lemire2020Chemical2}\\
  \ce{FeCl_4^-} & $197.657$ & $-536.870 \pm 3.072$ & 654.246 $\pm$ 11.44 & $144.922 \pm 64.30$ & & & \cite{Lemire2020Chemical2}\\
\\
  \textbf{Fe(II) carbonate complexes} \\
  \hline
  \ce{FeHCO_3^+} & $116.862$ & $-685.600$ & -776.3 & $36.550$ & $231.409$ & $0.818$ & \cite{Lemire2020Chemical2}\\
  \ce{FeCO3} (aq) & $115.854$ & $-648.677 \pm 1.285$ & -770.99 & $-62.000$ & $-123.025$ & $-1.723$ & \cite{Lemire2020Chemical2}\\
  \ce{Fe(CO_3)_2^2-} & $175.865$ & $-1186.67 \pm 1.293$ & -1384.72 & $128.000$ & & & \cite{Lemire2020Chemical2}\\
  \\
  \textbf{Solid Fe(II) hydroxides} \\
  \hline
  \ce{Fe(OH)2} (s) – White Rust & $89.8597$ & $-494.889 \pm 5.069$ & -583.39 & $84.00$ & $\approx 90.000$ & $2.952$ & \cite{Lemire2020Chemical2, Robie1995ThermodynamicTemperatures}\\
  \\
  \textbf{Solid Fe(III) hydroxide} \\
    \hline
  \ce{Fe(OH)3} (s) – 2-line ferrihydrite & $106.867$ & $-708.500 \pm 2.000$ & -827.100 $\pm$ 2.000 & $127.600 \pm 5.400$ & $\approx 152.000 \pm 5.00$ & $3.400$ & \cite{Dilnesa2011IronPhases}\\
  \ce{$\alpha$-FeOOH} (s) – Goethite & $88.8518$ & $-489.54 \pm 2.000$ & -560.50 $\pm$ 2.000 & $59.700 \pm 0.500$ & $ 74.400 \pm 0.40$ & $2.090$ & \cite{Lemire2020Chemical2}\\
  \\
  \textbf{Solid Fe(II) carbonates} \\
  \hline
  \ce{FeCO_3} (s) – Siderite & $71.8444$ & $-679.557 \pm 0.917$ & $-752.609 \pm 0.895$ & $95.537 \pm 0.646$ & $82.450 \pm 2.000$ & $2.938$ & \cite{Lemire2020Chemical2}\\
   \\
  \textbf{Calcium carbonate and related aqueous and solid species} \\
     \hline
  $\ce{CaCO3}$ (s) – Calcite & $100.086$ & $-1129.176$ & $-1207$ & $92.7$ & $81.9$ & & \cite{Lothenbach2019Cemdata18:Materials, Hummel2023The2020}\\
  $\ce{Ca(OH)2}$ (s) & $74.093$ & $-897.013$ & $-984.675$ & $83.399$ & $87.505$ & $3.306$ & \cite{Lothenbach2019Cemdata18:Materials, Hummel2023The2020}\\
  \ce{Ca^2+} & $40.078$ & $-552.79$ & $-543.07$ & $-56.48$ & $-30.92$ & $-1.844$ & \cite{Lothenbach2019Cemdata18:Materials, Hummel2023The2020}\\
  \ce{CaOH^+} & $57.085$ & $-717.02$ & $-751.65$ & $28.03$ & $6.05$ & $0.576$ & \cite{Lothenbach2019Cemdata18:Materials, Hummel2023The2020}\\
  \ce{CaCO3} (aq) & $100.086$ & $-1099.18$ & $-1201.92$ & $10.46$ & $-123.86$ & $-1.565$ & \cite{Lothenbach2019Cemdata18:Materials, Hummel2023The2020}\\
  \ce{CaHCO_3^+} & $101.094$ & $-1146.04$ & $-1231.94$ & $66.94$ & $233.70$ & $1.333$ & \cite{Lothenbach2019Cemdata18:Materials, Hummel2023The2020}\\
  \ce{CO_3^2-} & $60.009$ & $-527.900 \pm 0.390$ & $-675.230 \pm 0.250$ & $50.000 \pm 1.000$ & $-289.33$ & $-0.06$ & \cite{Lothenbach2019Cemdata18:Materials, Hummel2023The2020}\\
  \ce{HCO_3^-} & $61.017$ & $-586.845 \pm 0.251$ & $-689.930 \pm 0.200$ & $98.400 \pm 0.500$  & $-34.85$ & $2.421$ & \cite{Lothenbach2019Cemdata18:Materials, Hummel2023The2020}\\
  \ce{CO2} (aq) & $44.009$ & $-386.02$ & $-413.84$ & $117.57$ & $243.08$ & $3.281$ & \cite{Lothenbach2019Cemdata18:Materials, Hummel2023The2020}\\
  \\
   \textbf{Auxiliary species} \\
   \hline
  \ce{Na^+} & $22.9897$ & $-261.88$ & $-240.28$ & $58.41$ & $38.12$ & $-1.21$ & \cite{Lothenbach2019Cemdata18:Materials, Hummel2023The2020}\\
  \ce{Cl^-} & $35.453$ & $-131.29$ & $-167.11$ & $56.74$ & $-122.49$ & $17.34$ & \cite{Lothenbach2019Cemdata18:Materials, Hummel2023The2020}\\
  \ce{O2} (g) & $31.9988$ & $0$ & $0$ & $205.152 \pm 0.005$ & $29.378 \pm 0.003$ & $2478.970$ & \cite{Lothenbach2019Cemdata18:Materials, Hummel2023The2020}\\
  \ce{O2} (aq) & $31.9988$ & $16.45$ & $-12.24$ & $108.95$ & $234.13$ & $30.50$ & \cite{Lothenbach2019Cemdata18:Materials, Hummel2023The2020}\\
  \ce{H2} (g) & $2.0159$ & $0$ & $0$ & $13.680 \pm 0.003$ & $28.836 \pm 0.002$ & $2478.970$ & \cite{Lothenbach2019Cemdata18:Materials, Hummel2023The2020}\\
  \ce{H2} (aq) & $2.0159$ & $17.73$ & $-4.02$ & $57.74$ & $166.85$ & $25.26$ & \cite{Lothenbach2019Cemdata18:Materials, Hummel2023The2020}\\
  \ce{CO2} (g) & $44.009$ & $-394.39$ & $-393.51$ & $213.74$ & $37.15$ & & \cite{Lothenbach2019Cemdata18:Materials, Hummel2023The2020}\\
  \ce{H2O} (l) & $18.0153$ & $-237.140 \pm 0.041$ & $-285.830 \pm 0.040$ & $69.650 \pm 0.030$ & $75.351 \pm 0.080$ & $1.807$ & \cite{Lothenbach2019Cemdata18:Materials, Hummel2023The2020}\\
  \ce{H2O} (g) & $18.0153$ & $-228.582 \pm 0.040$ & $-241.826 \pm 0.040$ & $188.835 \pm 0.010$ & $33.609 \pm 0.030$ & $2478.970$ & \cite{Lothenbach2019Cemdata18:Materials, Hummel2023The2020}\\
  \ce{H^+} & $1.008$ & $0$ & $0$ & $0$ & $0$ & $0$ & \cite{Lothenbach2019Cemdata18:Materials, Hummel2023The2020}\\
  $\ce{OH^-}$ & $17.007$ & $-157.220$ & $-230.015 \pm 0.040$ & $-10.900 \pm 0.200$ & $-136.34$ & $-4.71$ & \cite{Lothenbach2019Cemdata18:Materials, Hummel2023The2020}\\
   \\
   \hline
   {$^{\text{a}}$ Internally computed from $\Delta_rG^\circ_m = \sum_i \nu_i \Delta_fG^\circ_{m,i}$.}\\
\end{tabular}
\end{adjustbox}
\label{table:Fe-thermo}
\end{table}
\newpage
Thermodynamic data for the solid, aqueous and gaseous species considered in this study for the Mn-C-\ce{H_2O} system.

\begin{table}[H]
\begin{minipage}{\textwidth}
\begin{adjustbox}{width=\columnwidth,center}
\begin{tabular}{llllllll}
\hline
    \textbf{Species} & \textbf{MW} (g/mol)  & $\mathbf{\Delta_fG^\circ_m}$, $\si{\kilo\joule\per\mole}$ & $\mathbf{\Delta_fH^\circ_m}$, $\si{\kilo\joule\per\mole}$  & $\mathbf{S^\circ_m}$, $\si{\joule\per\mole\per\kelvin}$  & $\mathbf{C_{p,m}^\circ}$, $\si{\joule\per\mole\per\kelvin}$  & $\mathbf{V^\circ}$, $\si{\joule\per\bar}$ & \textbf{Ref.} \\
  \hline
  \ce{Mn^2+} & $54.938$ & $-230.538$ & $-221.328$ & $-67.781$ & $-16.552$ & $-1.753$ & \cite{Hummel2023The2020}\\
  \ce{Mn^3+} & $54.938$ & $-84.935$ & $-128.449$ & $-309.62$ & $-105.95$ & $-4.1827$ & \cite{Hummel2023The2020}\\
  \hline
  \\
  \textbf{Mn(II) hydrolysis products} \\
  \hline
  \ce{MnOH^{+}} & $71.945$ & $-407.273^\text{a}$ & $-447.025$ & $1.255$ & $36.637$ & $-1.207$ & \cite{Hummel2023The2020,Brown2016HydrolysisIons}\\
  \ce{Mn(OH)_{2} (aq)} & $88.95$ & $-575.8 \pm 1.2^\text{a}$ & $-675.0 \pm 2.6$ & $35.1 \pm 9.8$ & $$ & $$ & \cite{Hummel2023The2020,Brown2016HydrolysisIons}\\
  \ce{Mn(OH)_{3}^{-}} & $105.93$ & $-743.6 \pm 2.6^\text{a}$\ & $-906.4 \pm 3.1$ & $55 \pm 14$ & $$ & $$ & \cite{Hummel2023The2020,Brown2016HydrolysisIons}\\
  \ce{Mn(OH)_{4}^{2-}} & $122.967$ & $-901.2 \pm 2.3^\text{a}$ & $-1107.7 \pm 5.2$ & $142 \pm 19$ & $$ & $$ & \cite{Hummel2023The2020,Brown2016HydrolysisIons}\\
  \\
  \textbf{Mn(III) hydrolysis products} \\
  \hline
  \ce{MnOH^{2+}} & $71.945$ & $-324.4 \pm 1.0^\text{a}$ & $-262.90 \pm 5.5$ & $276 \pm 19$ & $$ & $$ & \cite{Hummel2023The2020, Brown2016HydrolysisIons}\\
  \ce{Mn(OH)_2^+} & $88.95$ & $-565.842^\text{a}$ & $$ & $$ & $$ & $$ & \cite{Hummel2023The2020}\\
  \\
  \textbf{Mn(II) chloride complexes} \\
  \hline
  \ce{MnCl^{+}} & $90.391$ & $-365.310^\text{a}$ & $-373.659$ & $50.208$ & $105.258$ & $0.663$ & \cite{Hummel2023The2020}\\
  \\
  \textbf{Mn(III) chloride complexes} \\
  \hline
  \ce{MnCl^{2+}} & $90.391$ & $-218.783^\text{a}$ & $$ & $$ & $$ & $$ & \cite{Hummel2023The2020}\\
  \\
  \textbf{Mn(II) carbonate complexes} \\
  \hline
  \ce{MnCO_{3} (aq)} & $114.947$ & $-786.489$ & $-893.148$ & $-12.259$ & $-107.127$ & $-1.366$ & \cite{Hummel2002Nagra/PSI01/01}\\
  \ce{MnHCO_{3}^{+}} & $115.954$ & $-828.609$ & $-918.253$ & $44.810$ & $266.159$ & $1.437$ & \cite{Hummel2002Nagra/PSI01/01}\\
  \\
  \textbf{Solid Mn(II) phases} \\
  \hline
  \ce{Mn(OH)_{2} (s)} & $88.95$ & $-615.7 \pm 0.8$ & $-695.4$ & $99.2$ & $$ & $$ & \cite{Hummel2023The2020}\\
  \ce{MnO (s)} & $70.937$ & $-362.9 \pm 0.5$ & $-385.22 \pm 0.5$ & $59.71 \pm 0.4$ & $44.762$ & $1.322$ & \cite{Hummel2023The2020}\\
  \ce{MnCO_3 (s)} & $114.947$ & $-818.13 \pm 0.55$ & $-891.91 \pm 0.52$ & $98.03 \pm 0.1$ & $81.533$ & $3.1075$ & \cite{Hummel2023The2020, Hummel2002Nagra/PSI01/01}\\
  \\
  \textbf{Solid Mn(III) phases} \\
  \hline
  \ce{MnOOH (s)} & $87.945$ & $-557.8 \pm 2.0$ & $-650.453$ & & $79.652$ & & \cite{Hummel2023The2020}\\
  \\
  \textbf{Calcium carbonate and related species} \\
     \hline
  $\ce{CaCO3}$ (s) – Calcite & $100.086$ & $-1129.176$ & $-1207$ & $92.7$ & $81.9$ & & \cite{Lothenbach2019Cemdata18:Materials, Hummel2023The2020}\\
  $\ce{Ca(OH)2}$ (s) & $74.093$ & $-897.013$ & $-984.675$ & $83.399$ & $87.505$ & $3.306$ & \cite{Lothenbach2019Cemdata18:Materials, Hummel2023The2020}\\
  \ce{Ca^2+} & $40.078$ & $-552.79$ & $-543.07$ & $-56.48$ & $-30.92$ & $-1.844$ & \cite{Lothenbach2019Cemdata18:Materials, Hummel2023The2020}\\
  \ce{CaOH^+} & $57.085$ & $-717.02$ & $-751.65$ & $28.03$ & $6.05$ & $0.576$ & \cite{Lothenbach2019Cemdata18:Materials, Hummel2023The2020}\\
  \ce{CaCO3} (aq) & $100.086$ & $-1099.18$ & $-1201.92$ & $10.46$ & $-123.86$ & $-1.565$ & \cite{Lothenbach2019Cemdata18:Materials, Hummel2023The2020}\\
  \ce{CaHCO_3^+} & $101.094$ & $-1146.04$ & $-1231.94$ & $66.94$ & $233.70$ & $1.333$ & \cite{Lothenbach2019Cemdata18:Materials, Hummel2023The2020}\\
  \ce{CO_3^2-} & $60.009$ & $-527.900 \pm 0.390$ & $-675.230 \pm 0.250$ & $50.000 \pm 1.000$ & $-289.33$ & $-0.06$ & \cite{Lothenbach2019Cemdata18:Materials, Hummel2023The2020}\\
  \ce{HCO_3^-} & $61.017$ & $-586.845 \pm 0.251$ & $-689.930 \pm 0.200$ & $98.400 \pm 0.500$  & $-34.85$ & $2.421$ & \cite{Lothenbach2019Cemdata18:Materials, Hummel2023The2020}\\
  \ce{CO2} (aq) & $44.009$ & $-386.02$ & $-413.84$ & $117.57$ & $243.08$ & $3.281$ & \cite{Lothenbach2019Cemdata18:Materials, Hummel2023The2020}\\
  \\
   \textbf{Auxiliary species} \\
   \hline
  \ce{Na^+} & $22.9897$ & $-261.88$ & $-240.28$ & $58.41$ & $38.12$ & $-1.21$ & \cite{Lothenbach2019Cemdata18:Materials, Hummel2023The2020}\\
  \ce{Cl^-} & $35.453$ & $-131.29$ & $-167.11$ & $56.74$ & $-122.49$ & $17.34$ & \cite{Lothenbach2019Cemdata18:Materials, Hummel2023The2020}\\
  \ce{O2} (g) & $31.9988$ & $0$ & $0$ & $205.152 \pm 0.005$ & $29.378 \pm 0.003$ & $2478.970$ & \cite{Lothenbach2019Cemdata18:Materials, Hummel2023The2020}\\
  \ce{O2} (aq) & $31.9988$ & $16.45$ & $-12.24$ & $108.95$ & $234.13$ & $30.50$ & \cite{Lothenbach2019Cemdata18:Materials, Hummel2023The2020}\\
  \ce{H2} (g) & $2.0159$ & $0$ & $0$ & $13.680 \pm 0.003$ & $28.836 \pm 0.002$ & $2478.970$ & \cite{Lothenbach2019Cemdata18:Materials, Hummel2023The2020}\\
  \ce{H2} (aq) & $2.0159$ & $17.73$ & $-4.02$ & $57.74$ & $166.85$ & $25.26$ & \cite{Lothenbach2019Cemdata18:Materials, Hummel2023The2020}\\
  \ce{CO2} (g) & $44.009$ & $-394.39$ & $-393.51$ & $213.74$ & $37.15$ & & \cite{Lothenbach2019Cemdata18:Materials, Hummel2023The2020}\\
  \ce{H2O} (l) & $18.0153$ & $-237.140 \pm 0.041$ & $-285.830 \pm 0.040$ & $69.650 \pm 0.030$ & $75.351 \pm 0.080$ & $1.807$ & \cite{Lothenbach2019Cemdata18:Materials, Hummel2023The2020}\\
  \ce{H2O} (g) & $18.0153$ & $-228.582 \pm 0.040$ & $-241.826 \pm 0.040$ & $188.835 \pm 0.010$ & $33.609 \pm 0.030$ & $2478.970$ & \cite{Lothenbach2019Cemdata18:Materials, Hummel2023The2020}\\
  \ce{H^+} & $1.008$ & $0$ & $0$ & $0$ & $0$ & $0$ & \cite{Lothenbach2019Cemdata18:Materials, Hummel2023The2020}\\
  $\ce{OH^-}$ & $17.007$ & $-157.220$ & $-230.015 \pm 0.040$ & $-10.900 \pm 0.200$ & $-136.34$ & $-4.71$ & \cite{Lothenbach2019Cemdata18:Materials, Hummel2023The2020}\\
   \\
  \hline
  {$^{\text{a}}$ Internally computed from $\Delta_rG^\circ_m = \sum_i \nu_i \Delta_fG^\circ_{m,i}$.}\\
    \end{tabular}
    \label{table:Mn-thermo}
\end{adjustbox}
\end{minipage}
\end{table}

\newpage
\section{Schematic representation of operator-split thermodynamic modelling}\label{sec:SM3-staggeredthermo}

\begin{figure}[H]
    \centering
    \includegraphics[width=0.75\linewidth]{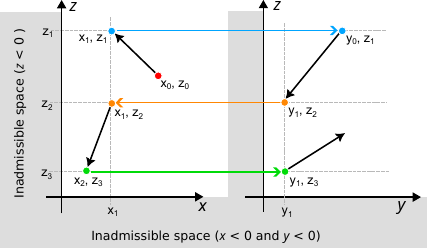}
    \caption{Schematic representation of operator-split algorithm for thermodynamic modelling of multiple redox states. Here $x$ refers to the total concentration of all $\metal^\astate$ complexes, $y$ refers to the total concentration of all $\metal^\bstate$ complexes and $z$ refers to the total concentration of all the background species. For visual representation, we show the concentration of all background species as one axis, however, in reality the space is multi-dimensional with individual axes for each background species}
    \label{fig:schematic-thermo}
\end{figure}

\Cref{fig:schematic-thermo} shows the schematic representation of the operator-split algorithm for thermodynamic modelling. Although, the algorithm works in a $n$-dimensional space (where $n$ is the total number of species in the chemical system), here for ease of understanding we have reduced it to a 2-dimensional space.

\newpage
\section{Validation of time-step for SNIA aprroach}\label{sec:SM4-SNIA}

In our reactive transport framework, we employ a staggered non-iterative approach (SNIA) to couple the transient processes such as diffusion and kinetically controlled reactions with the non-transient processes such as chemical reactions at equilibrium. We performed a sensitivity analysis to find the optimal time step for temporal evolution. \Cref{fig:time-step} shows the effect of time step on the spatial evolution of the concentration of the chemical species. We note that we hose the time step value for which the results are converged. 
\begin{figure}[H]
    \centering
    \includegraphics[width=0.9\textwidth]{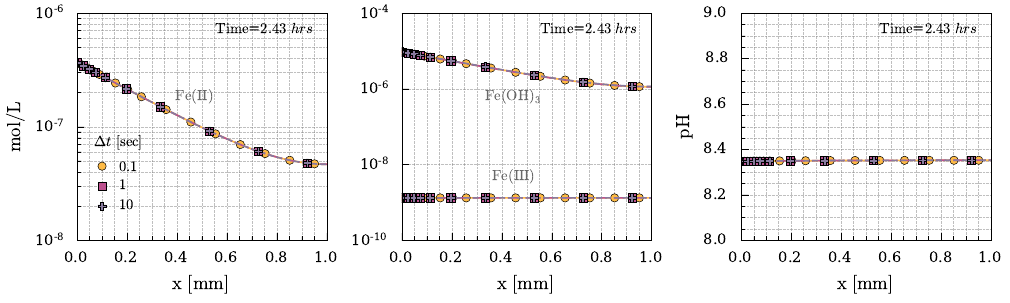}
    \caption{Sensitivity analysis to investigate the effect of time step (0.1 sec to 10 sec) on the modelling of reactive transport using an SNIA approach. }
    \label{fig:time-step}
\end{figure}

\newpage
\section{Processes that can influence the interfacial chemistry in the context of steel corrosion in concrete}\label{sec:SM5-corrosion}

The pore solution of the concrete is generally highly alkaline (pH: $>$ 12.5), facilitating the development of a robust passive film on the steel. The breakdown of steel passivity in concrete can be either local due to the ingress of chlorides or uniform due to the lowering of the pH of the pore solution due to carbonation. When considering localised corrosion, the breakdown of passivity may cause a change in the local pH in the area of active corrosion. However, this change in the local pH might be buffered for a significantly long time due to the dissolution of \ce{Ca(OH)_2} (s) present in cementitious matrix. The carbonation of the concrete cover (i.e. decalcification of Ca-bearing cement hydrate phases and the precipitation of \ce{CaCO3}) and the associated loss of alkalinity (pH $<$ 9) is often 'perceived' as a cause for the corrosion of the embedded steel reinforcement \cite{vonGreve-Dierfeld2020Understanding281-CCC}. However, in the context of infrastructure corrosion, steel corrosion rates are generally only of concern when the internal relative humidity at the steel-concrete interface (SCI) is $\geq$ 80 – 95 \% ~\cite{Stefanoni2020TheExposure, Angst2020}. In the case of fully saturated and completely carbonated concretes, the corrosion of the embedded reinforcement may be sustained. 

The sustained anodic dissolution of iron leads to the production of \ce{Fe^2+} (\Cref{eq:S1}) and the simultaneous cathodic generation of \ce{OH^-} (the flux of [\ce{OH^-}] is twice that of [\ce{Fe^2+}]). In the aqueous phase, \ce{Fe^2+} may hydrolyse (\Cref{eq:S2}) (depending on the pH) and may oxidise to \ce{Fe^3+} or any of the \ce{Fe^{III}} hydrolysis products (\Cref{eq:S3} and \Cref{eq:S4}), releasing protons. Upon reaching the solubility limit w.r.t. to different iron (hydr)oxides, both \ce{Fe^{II}} and \ce{Fe^{III}} can undergo precipitation as represented for the simple cases of \ce{Fe(OH)_2} and \ce{Fe(OH)_3} by \Cref{eq:S5} and \Cref{eq:S6}, respectively, all causing changes in the chemistry of the aqueous phase. These changes in the aqueous chemistry influence the kinetics of oxidation, hydrolysis, speciation and precipitation of Fe. The ever-evolving chemistry may influence the rate at which the metal (iron in this case) corrodes. 

\begin{equation}\label{eq:S1}
    \ce{Fe -> Fe^{2+} +2e-}
\end{equation}
\begin{equation}\label{eq:S2}
    \ce{Fe^{2+} + mH2O -> Fe(OH)_{m}^{2-m} + mH+}
\end{equation}
\begin{equation}\label{eq:S3}
    \ce{Fe(II) -> Fe(III) + e-}
\end{equation}
\begin{equation}\label{eq:S4}
    \ce{Fe^{3+} + nH2O -> Fe(OH)_{n}^{3-n} + nH+}
\end{equation}
\begin{equation}\label{eq:S5}
    \ce{Fe^{2+} + 2H2O -> Fe(OH)_{2} (s) + 2H+}
\end{equation}
\begin{equation}\label{eq:S6}
    \ce{Fe^{3+} + 3H2O -> Fe(OH)_{3} (s) + 3H+}
\end{equation}

In fully carbonated concrete, the pore solution is in equilibrium with \ce{CaCO_3} (s) surrounding the pore network. Changes in the pH of the pore solution due to the reactions mentioned in \Cref{eq:S1} - \Cref{eq:S6}, may be countered/buffered by the dissolution of \ce{CaCO_3} (s).  

\subsection{Acid-neutralising capacity as a function of [\ce{CaCO_3}] (s)}\label{SM5.1-CaCO3}

The degree of super/under saturation w.r.t. \ce{CaCO_3} (s) influences the rate of change of pH, speciation of aqueous carbonate and the concentration of mobile metal cations. \Cref{fig:calcium-solubility}a and \Cref{fig:calcium-solubility}b show the influence of [\ce{CaCO_3}] in anaerobic and aerobic environments, for the example of steel corrosion in concrete. The \ce{CaCO_3} (s) can dissolve to produce \ce{HCO_3^-} and \ce{CO_3^{2-}} to buffer any changes in the pH. Therefore, the acid neutralising capacity of the interfacial electrolyte is strongly dependent on the degree of super/under saturation w.r.t. \ce{CaCO_3}; and a higher [\ce{CaCO_3}] would yield a lower rate of acidification \Cref{fig:Influence of diffusivity}e). This is analogous to the influence of anthropogenic \ce{CO_2} on the dissolution of \ce{CaCO_3} in the oceans, where the degree of supersaturation of \ce{CaCO_3} as a function of the oceanic depth controls the acid neutralising capacity of the waters ~\cite{Sulpis2018CurrentCO2, Schnoor1986TheDeposition, Sulpis2021CalciumOcean, Feely2004ImpactOceans, Morse2007CalciumDissolution, Morse2006InitialMg-calcites, Stumm1996AquaticWaters}.

\begin{figure}[H] 
    \centering
    \includegraphics[width=0.95\textwidth]{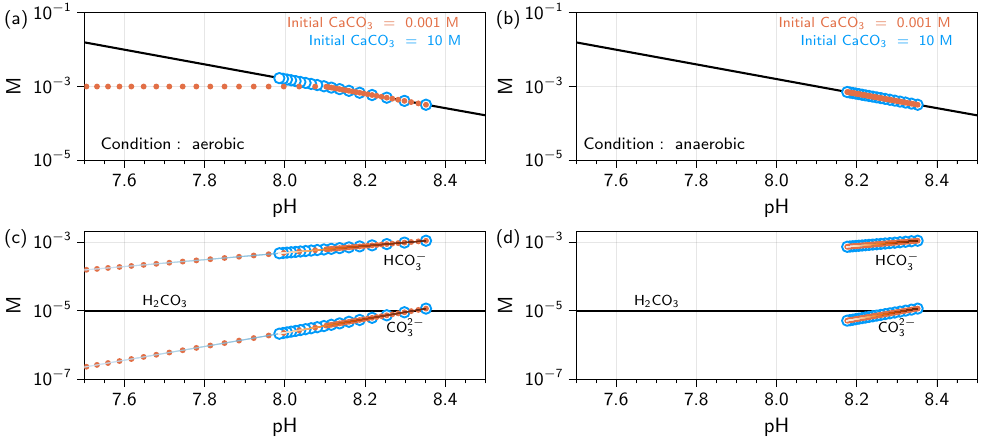}
    \caption{The acid-neutralising capacity of \ce{CaCO_3} is shown through the solubility of \ce{CaCO_3} (s) and speciation of aqueous carbonate species under aerobic and anaerobic conditions. The blue and red markers indicate the pH and the corresponding speciation of aqueous carbonate species when the initial [\ce{CaCO_3}] is 10 mol/L and 0.001 mol/L, respectively}
    \label{fig:calcium-solubility}
\end{figure} 

\subsection{Influence of [\ce{CaCO_3}] and oxygen on pH buffering}
Let us look at the influence of [\ce{CaCO_3}] surrounding the pore network on the pH buffering at the metal-porous medium interface. We assume that \ce{CaCO_3}, formed as a result of the carbonation of the binder, can instantaneously dissolve into bicarbonate and carbonate anions, and their respective aqueous molar fractions depend on the prevailing pH. Therefore, we do not consider the kinetics of \ce{CaCO_3} dissolution as a function of the gradients in the pH, and speciation of aqueous inorganic carbon at the surface of the dissolving \ce{CaCO_3} (s). Regardless of whether the interfacial electrolyte is oxygenated or deoxygenated, the higher the amount of \ce{CaCO_3}, the lower the rate at which the interfacial pH decreases (as seen in \Cref{fig:oxygen-interface}a, d). Over the course of 72 hours in aerobic and anaerobic electrolytes and when 10 mol/L \ce{CaCO3} surrounds the interfacial electrolyte, $\mathrm{\Delta~pH}$ at the interface is $\approx$ -0.4 and -0.25, respectively (\Cref{fig:oxygen-interface}a, d). Whereas for [\ce{CaCO3}] = \ce{10^-3} mol/L, $\mathrm{\Delta~pH}$ at the interface over 72 hours is $\approx$ -1.1 and -0.25, for aerobic and anaerobic conditions, respectively (\Cref{fig:oxygen-interface}a, d). In the absence of \ce{CaCO_3}, the pH reduces by approximately 2 units in aerobic conditions and 0.8 units in anaerobic conditions, over the same duration (\Cref{fig:oxygen-interface}a, d). Upon a decrease in the interfacial pH (due to corrosion of steel and associated process such as hydrolysis, oxidation and precipitation, and continuous exposure to atmospheric \ce{CO_2} (g)), the \ce{CaCO_3} surrounding the pore network dissolves and produces \ce{HCO_3^-} to neutralise the interfacial solution. We further highlight the role of degree of supersaturation w.r.t. \ce{CaCO_3} by observing the reduction in the pH in aerobic interfacial electrolytes surrounded by \ce{10^{-3}} mol/L \ce{CaCO_3}. The pH decreases slowly from 8.4 to 8.05, in the same way and rate as in the case of [\ce{CaCO_3}] = 10 mol/L until $\approx$ 36 hours (\Cref{fig:oxygen-interface}a, d). However, upon reaching undersaturation w.r.t. \ce{CaCO_3} at 36 hours, the acid neutralising capacity of the interfacial solution reduces further, and the pH decreases drastically by almost 1 unit over the next 36 hours (\Cref{fig:calcium-solubility}).

\begin{figure}[!ht]
    \centering
    \includegraphics[width=0.95\textwidth]{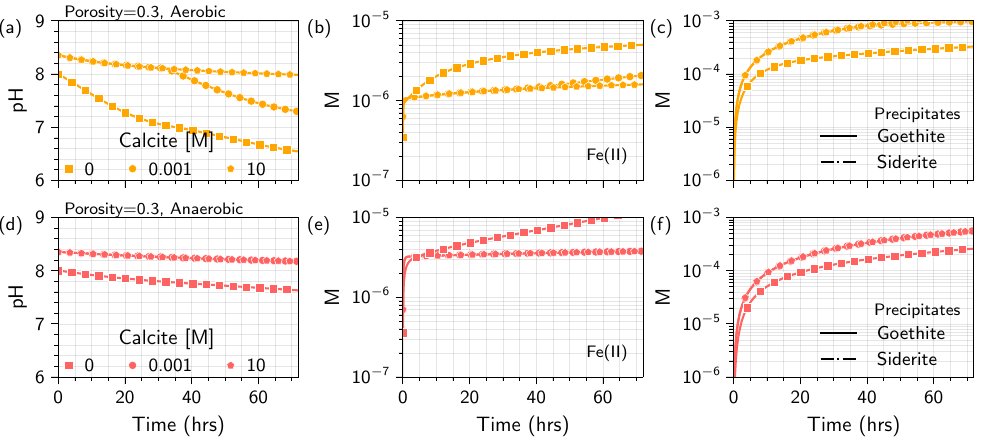}
    \caption{\textbf{Influence of oxygen (aerobic vs. anaerobic) and [\ce{CaCO_3}] on the chemistry at the metal-porous medium interface:} The results are shown for a region at a distance $\sim 0.0025$ mm from steel-porous medium interface. We consider 3 different starting concentrations (10 mol/l, 0.001 mol and 0 mol) of \ce{CaCO_3} and for a fixed fugacity of \ce{CO_2} at $10^{-3.5}$ atm. The one-dimensional porous medium has a porosity of 0.3. (a)-(c) Evolution of pH, total ferrous complexes and precipitates respectively for aerobic conditions. (d)-(f) Evolution of pH, total ferrous complexes and precipitates respectively for anaerobic conditions}
    \label{fig:oxygen-interface}
\end{figure}

The activity of dissolved oxygen, a very relevant parameter in electrochemical studies, has a significant influence on the chemistry of the interface. The decrease in the interfacial pH is more rapid and pronounced in aerobic environments (\Cref{fig:Influence of diffusivity}i) than anaerobic conditions (\Cref{fig:Influence of diffusivity}l). The higher rate of \ferrous{} oxidation (by $\approx$ 2 orders of magnitude in aerobic conditions than anaerobic conditions over the pH of interest - \Cref{fig:supp-fe}c) in aerobic conditions is responsible for a higher concentration of \ferric{} produced over a given time interval and therefore, the amount of goethite precipitated. Thus, the excess protons produced as a result of precipitation (\Cref{sec:SM5-corrosion}) are much higher in aerobic conditions than anaerobic conditions and the decrease in the interfacial pH over time is higher (compare \Cref{fig:Influence of diffusivity}i and \Cref{fig:Influence of diffusivity}l). 

The influence of oxygen on the interfacial chemistry is particularly evident in the absence of \ce{CaCO_3} and when the [\ce{CaCO_3}] = \ce{10^-3} mol/L. The lower rate of \ce{Fe^{II}} oxidation (by $\approx$ 2 orders of magnitude in anaerobic conditions than aerobic conditions over the pH of interest - \Cref{fig:supp-fe}c) under anaerobic conditions is responsible for a lower concentration of \ce{Fe^{II}} produced over a given time interval and therefore, the amount of goethite precipitated. Thus, the excess protons produced as a result of precipitation are much lower in anaerobic conditions than aerobic conditions and the decrease in the interfacial pH over time is lower. Intuitively, we expect the lower pH at the interface to facilitate an increase in the solubility of \ce{Fe^{II}} \cite{furcas_solubility_2022} and also a decrease in the\ce{Fe^{II}} oxidation rate, resulting in increased [\ce{Fe^{II}}] over time. This argument is generally true and can explain the evolution of [\ce{Fe^{II}}] over time in individual scenarios for varying [\ce{CaCO3}] and a particular oxygen concentration. However, upon comparing aerobic and anaerobic conditions, the [\ce{Fe^{II}}] is significantly higher in the latter case where the pH is relatively higher at all times. The primary reason for this is the slower rate of \ce{Fe^{II}} oxidation under anaerobic conditions~\cite{Mundra2023AerobicSolutions}, which decreases the rate at which \ce{Fe^{II}} is consumed. 

\subsection{Solubility of \ferrous{} in the presence of \ce{CaCO_3}}\label{sec:si-ferrous-stability}
The solubility of \ce{Fe^{II}} is strongly influenced by the [\ce{CaCO_3}]. As the concentration of \ce{CaCO_3} increases from \ce{10^{-4} mol/L} to 10 mol/L, the solubility of \ce{Fe^{II}} increasingly depends on the aqueous iron-carbonate complexes, particularly in the near neutral pH range. \Cref{fig:ferrous-solubility} indicates the overall solubility of \ce{Fe^{II}} and the contributions of each of the species formed in the \ce{Fe^{II}}-\ce{CaCO_3}-\ce{H_2O} system as a function of the pH. It is worthwhile to note that in the presence of higher [\ce{CaCO_3}], the overall solubility of \ce{Fe^{II}} is controlled by \ce{FeCO_3} (s), particularly at near neutral pH conditions (6 $<$ pH $<$ 10). However, \ce{Fe(OH)_2} (s) controls the solubility of \ce{Fe^{II}} in alkaline conditions (pH $>$ 10), irrespective of the [\ce{CaCO_3}].

\begin{figure}[H]
    \centering
    \includegraphics[width=0.95\textwidth]{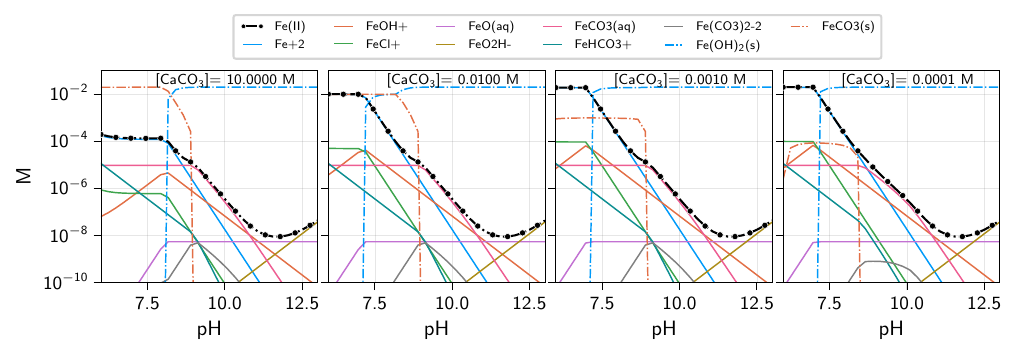}
    \caption{Solubility of \ferrous{} under aerobic conditions in the presence of different concentrations of \ce{[CaCO_3]}. A [\ce{Fe^2+}] of 0.02 mol/L was used to compute the solubility of \ferrous{} species.}
    \label{fig:ferrous-solubility}
\end{figure}

\subsection{\ferric{} at the metal-porous medium interface and further away from the metal surface}

\begin{figure}[H]
    \centering
    \includegraphics[width=0.95\textwidth]{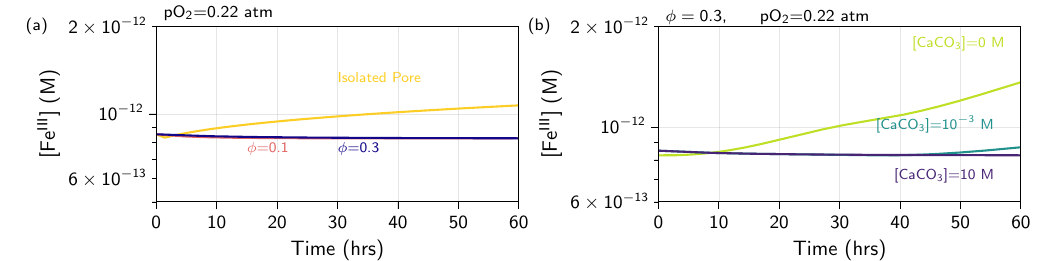}
    \caption{Influence of average porosity (a) and the [\ce{CaCO_3}] (b) on the concentration of \ferric{} over time at the metal-porous medium interface}
    \label{fig:ferric-interface-conc-speciation}
\end{figure}

\begin{figure}[H]
    \centering
    \includegraphics[width=0.95\textwidth]{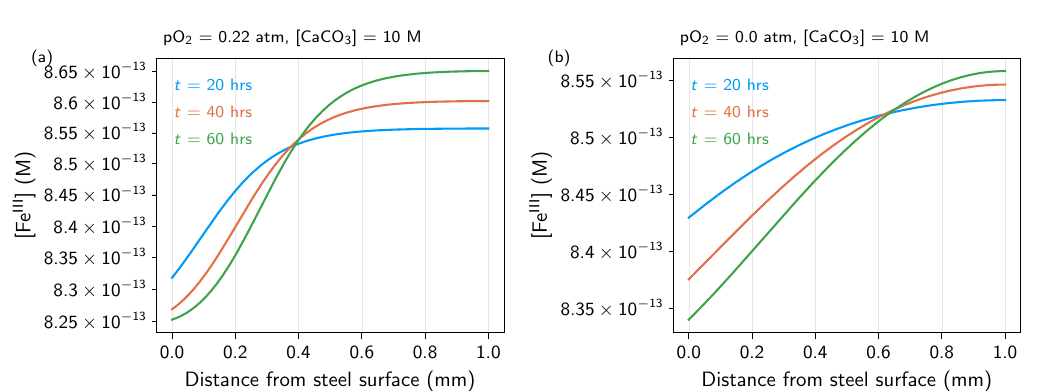}
    \caption{Spatial and temporal distribution of [\ferric{}] in aerobic (a) and anaerobic (b) environments within the porous medium}
    \label{fig:ferric-conc-space}
\end{figure}

\subsection{Framework limitations}
First, we assume that \ce{Fe^{III}} formed as a result of the oxidation of \ce{Fe^{II}} ions instantly precipitate to its thermodynamically favoured solid state - goethite, and aqueous [\ce{Fe^{III}}] is dictated by the solubility limit of goethite (\Cref{fig:ferric-conc-space} and \Cref{fig:ferric-interface-conc-speciation}). It has been recently shown that thermodynamically less stable phases such as ferrihydrite may also precipitate~\cite{Albert2024MicroscaleInterface} in the porous medium. It is certainly the case that thermodynamically less stable \ce{Fe^{III}} (hydr)oxides (such as ferrihydrite) may transform to thermodynamically more stable states with time, which is kinetically controlled by the dissolution rate of the preceding solid phase~\cite{Furcas2023TransformationPH}. As a first approximation, here, we do not consider the time-dependent transformation rate of different solid \ce{Fe^{III}} (hydr)oxides. Furthermore, also the precipitation of corrosion products with iron in different oxidation states, such as magnetite (\ce{Fe_3O_4}), has not been considered in this work but, will be addressed in detail in future studies (as highlighted in \Cref{sec-outlook}). 

Aqueous \ce{Fe^{III}} is assumed to sorb on the surfaces of solid phases and cannot diffuse away. This assumption may not be entirely true. Therefore, in addition to the precipitation of \ce{Fe^{III}} hydr(oxides), another sink term taking into account \ce{Fe^{III}} sorption must be considered within the transient processes in our framework. 

A second limitation of our study pertains to the reactive solid phase, \ce{CaCO_3}, surrounding the electrolyte in the pore network. While we allow for the congruent dissolution of \ce{CaCO_3}, we do not consider the kinetics of \ce{CaCO_3} dissolution as a function of the gradients in the pH, and speciation of aqueous inorganic carbon at the surface of the dissolving \ce{CaCO_3} solid. To incorporate this, one would have to simulate processes occurring at the pore-scale within both the transient and equilibrium stages of this framework.

Thirdly, the precipitation of solid corrosion products, with a relatively higher molar volume compared to the bare metal within the pore network will change the local porosity and diffusivity of individual pores. It will also result in an increase in the expansive stresses exerted on the surrounding solids leading to the development of micro-cracks and subsequently altering the local pore structure. While, this is not the focus of this manuscript, one could incorporate temporal and spatial changes in the pore structure into the framework.

Finally, in the relevant example of steel corrosion in carbonated concrete and to illustrate the applicability of our reactive transport framework, we did not alter the \ce{Fe^{0}} corrosion rate over time and instead, used a constant flux of \ce{Fe^{II}} and correspondingly, [\ce{OH^-}]. \citet{Stefanoni2019KineticsMedia} established that the corrosion rate, under practical exposure conditions depends on the physical and chemical characteristics (fraction of saturated pores at the steel/porous media interface, the concentration of iron ions at the interface, diffusion of iron ions away from the corroding interface - dependent on porosity and connectivity) of the steel/porous media interface. In future studies we aim to incorporate this by changing the input flux accordingly in our framework.

\else
\fi

\end{document}